\documentclass[11pt]{article}
\usepackage{amsfonts,amssymb}
\usepackage{color}
\usepackage{pstricks,pst-node}
\usepackage{pst-plot}
\catcode`\@=11
\def\marginnote#1{}

\newcount\hour
\newcount\minute
\newtoks\amorpm
\hour=\time\divide\hour by60 \minute=\time{\multiply\hour by60
\global\advance\minute by-\hour}
\edef\standardtime{{\ifnum\hour<12 \global\amorpm={am}%
        \else\global\amorpm={pm}\advance\hour by-12 \fi
        \ifnum\hour=0 \hour=12 \fi
        \number\hour:\ifnum\minute<10 0\fi\number\minute\the\amorpm}}
\edef\militarytime{\number\hour:\ifnum\minute<10 0\fi\number\minute}

\def\appendix#1{
\addtocounter{section}{1} \setcounter{equation}{0}
\renewcommand{\thesection}{\Alph{section}}
\section*{Appendix \thesection\protect\indent\quad
#1}
}
\renewcommand{\theequation}{\thesection.\arabic{equation}}

\def\draftlabel#1{{\@bsphack\if@filesw {\let\thepage\relax
      \xdef\@gtempa{\write\@auxout{\string
          \newlabel{#1}{{\@currentlabel}{\thepage}}}}}\@gtempa \if@nobreak
    \ifvmode\nobreak\fi\fi\fi\@esphack} \gdef\@eqnlabel{#1}}
    \def\@eqnlabel{}
\def\@vacuum{}
\def\draftmarginnote#1{\marginpar{\raggedright\scriptsize\tt#1}}

\def\draft{
  \oddsidemargin -.5truein
  \def\@oddfoot{\footnotesize \sl preliminary draft \hfil
    \rm\thepage\hfil\sl\today\quad\militarytime}
  \let\@evenfoot\@oddfoot \overfullrule 3pt
    \let\label=\draftlabel
    \let\marginnote=\draftmarginnote
  \def\@eqnnum{(\theequation)\rlap{\kern\marginparsep\tt\@eqnlabel}%
    \global\let\@eqnlabel\@vacuum}

  }

\voffset= - 0.5in \hoffset= - 1.0in

\def\be{\begin{equation}}
\def\ee{\end{equation}}
\def\bea{\begin{eqnarray}}
\def\eea{\end{eqnarray}}
\def\<{\langle}
\def\>{\rangle}

\def\tr{{\mathrm{tr\,}}}

\def\1N{${\cal N}=1$}
\def\4N{${\cal N}=4$}

\newcommand{\bp}{\mathbb{P}}
\newcommand{\bz}{\mathbb{Z}}
\newcommand{\modm}{\mathcal{M}}

\def\e{{\,\rm e}\,}

\def\bea{\begin{eqnarray}}
\def\eea{\end{eqnarray}}

\def\pa{\partial}

\def\beq{\begin{equation}}
\def\eeq{\end{equation}}
\def\ba{\beq\begin{array}{c}}
\def\ea{\end{array}\eeq}

\@ifundefined{theorem@style}{\RequirePackage{theorem}}{}

\gdef\th@plain{\normalfont\slshape
  \def\@begintheorem##1##2{%
\item[\hskip\parindent\hskip\labelsep\theorem@headerfont ##1\ ##2\unskip.]}%
\def\@opargbegintheorem##1##2##3{%
\item[\hskip\parindent
\ifx\empty##1\else\hskip\labelsep\fi\theorem@headerfont ##1\ ##2\unskip]{\theorem@headerfont{\rm ##3}.} }}

\gdef\th@definition{\normalfont
  \def\@begintheorem##1##2{%
\item[\hskip\parindent\hskip\labelsep\theorem@headerfont ##1\ ##2\unskip.]}%
\def\@opargbegintheorem##1##2##3{%
\item[\hskip\parindent
\ifx\empty##1\else\hskip\labelsep\fi\theorem@headerfont ##1\ ##2\unskip]{\theorem@headerfont{\rm ##3}.} }}

\theoremstyle{plain}
\newtheorem{theorem}{Theorem}
\newtheorem{lemma}{Lemma}
\newtheorem{corollary}{Corollary}
\newtheorem{proposition}{Proposition}
\newtheorem{conjecture}{Conjecture}

\theoremstyle{definition}

\newtheorem{remark}{Remark}
\newtheorem{example}{Example}

\let\operatorname=\mathrm
\let\text=\mathrm

\let\wtd=\widetilde
\let\wht=\widehat

\def\ln{\operatorname{log}}

\def\Aut{\operatorname{Aut}}

\def\disc{\text{disc}}

\def\beq{\begin{equation}}
\def\eeq{\end{equation}}
\def\bea{\begin{eqnarray}}
\def\eea{\end{eqnarray}}

\newcommand{\binom}[2]{\left({#1\atop #2}\right)}

\newcommand{\cpict}[3]{
\dimen1=#1\advance\dimen1 by-\hsize\divide\dimen1 by-2 \vtop to #2{
\noindent\hskip\dimen1{\special{em:graph #3.bmp}} \vfil}\hskip-2cm }

\let\@@savethanks\thanks
\def\thanks#1{\gdef\thefootnote{\alph{footnote}}\@@savethanks{#1}}

\newcommand{\cl}{\mathcal{L}}
\newcommand{\bc}{\mathbb{C}}

\newcommand {\dvol}{discrete volumes}

\makeatletter
\def\blfootnote{\xdef\@thefnmark{}\@footnotetext}
\makeatother

\begin{document}

\title{Models of discretized moduli spaces, cohomological field theories,\\ and Gaussian means}

\author{J{\o}rgen Ellegaard Andersen\thanks{\noindent QGM, {\AA}rhus University, Denmark and Caltech, Pasadena, USA},
Leonid O. Chekhov\thanks{Steklov Mathematical Institute and  Laboratoire Poncelet,
Moscow, Russia, and QGM, {\AA}rhus University, Denmark},
Paul Norbury\thanks{University of Melbourne, Australia},
and
Robert C. Penner\thanks{IHES, Bures-sur-Yvette, France, and Caltech, Pasadena, USA}}


\maketitle

\begin{abstract}
We prove combinatorially the explicit relation between genus filtrated $s$-loop means 
of the Gaussian matrix model and terms of the genus expansion of the 
Kontsevich--Penner matrix model (KPMM). The latter is the
generating function for volumes of discretized (open) moduli spaces
$M_{g,s}^{\disc}$ given by $N_{g,s}(P_1,\dots,P_s)$ for $(P_1,\dots,P_s)\in\bz_+^s$. This generating function therefore enjoys the topological recursion, and we prove that it is 
simultaneously the generating function
for ancestor invariants of a 
cohomological field theory thus enjoying the Givental decomposition. We use another
Givental-type decomposition obtained for this model by the second authors in 1995 in terms of
special times related to the discretisation of moduli spaces thus representing its asymptotic
expansion terms (and therefore those of the Gaussian means) as finite sums over
graphs weighted by lower-order monomials in times thus giving another proof of (quasi)polynomiality of the \dvol.
As an application, we find the coefficients in the first subleading order for ${\mathcal M}_{g,1}$ in two ways:
by using the refined Harer--Zagier recursion and by exploiting the above Givental-type transformation. 
We put forward the conjecture that the above graph expansions can be used for probing the reduction structure of
the Delgne--Mumford compactification
 $\overline{\mathcal M}_{g,s}$ of moduli spaces of punctured Riemann surfaces.
\end{abstract}

\section{Introduction}

After Harer and Zagier derived~\cite{HZ} the celebrated recursion formula describing means of traces $\<\tr H^k\>$ of
powers of the $N\times N$-Hermitian matrix $H$ over the Gaussian ensemble, progress in this direction was mainly due to explicit (commonly multiple-integral) formulas for the generation functions for multi-trace means of the form
$\<\prod_{i=1}^s \tr H^{k_i}\>^{{\mathrm{conn}}}$. Brez\'in and Hikami~\cite{BreHik} used the replica
method ameliorated in~\cite{MS10} for obtaining {\em exact} $s$-fold integral representation for these  quantities valid for all $N$.  There has been revived interest in the multi-trace means due to their relation to {\em topological recursion} \cite{ChEy,EOrInv} and quantum curves \cite{GSuApo,MSuSpe,DM14}.  One aim of the present paper is to describe consequences of these recent ideas.  In particular, the relation to the Euler characteristic of the moduli space of curves, which is the main application in \cite{HZ}, is shown here to go much deeper and culminates in a {\em cohomological field theory} \cite{ManFro} equivalent to the multi-trace Gaussian means. 

In this paper we exploit the interplay between three main tools for studying the multi-trace Gaussian means: 
\begin{enumerate}
\item A matrix model, known as the Kontsevich--Penner matrix model (KPMM), with external field that describes all multi-trace Gaussian means, introduced in~\cite{ChMak1} and developed further in~\cite{ACKM}. 
\item Exact formulae, or equivalently integrals over $N\times N$ Hermitian matrices for fixed $N$, in place of asymptotic expansions of means in $N$ which gives rise to surfaces and their genus. 
\item Topological recursion satisfied by resolvents storing the multi-trace Gaussian means, which leads to a cohomological field theory with primary correlators given by $\chi(\modm_{g,s})$.
\end{enumerate}

The construction of the Kontsevich--Penner matrix model uses a standard combinatorial approach.  As was shown in~\cite{ChMak2} using the
Virasoro constraints and in~\cite{MMM} using the direct determinant relations, this model is
equivalent to the Hermitian matrix model with the potential whose times (coupling constants) are related to
the external-matrix eigenvalues via the Miwa-type transformation and whose matrix size is the coefficient of
the logarithmic term.  More precisely, the statement, which appears here as Corollary~\ref{cor:loopmean}, is that the 
free energy for the KPMM is a primitive (antiderivative) for the resolvents of the Gaussian matrix model.  The resolvents storing the multi-trace Gaussian means are naturally described as meromorphic (multi)differentials with zero residues over a rational Riemann surface, known as the {\em spectral curve}, hence their primitives are meromorphic functions on the spectral curve.  These primitives are conjecturally related (and proven in the Gaussian case \cite{MSuSpe}) to the so called {\em quantum curve} which is a linear differential equation that is a non-commutative quantisation of the spectral curve.  Corollary~\ref{cor:loopmean} can be used to prove the relationship between the spectral and quantum curves---it shows that the wave function arising out of the spectral curve is a specialization of the free energy for the KPMM which satisfies the second order differential equation that is the quantum curve.

The geometric content of the Kontsevich--Penner matrix model also proved
to be rather rich: its free energy was related to structures of {\em discretized moduli spaces} in~\cite{Ch93}; more
recently in ~\cite{NorStr} and \cite{Mulase-Penkawa}  it was identified with the generating function for \dvol\ 
$N_{g,s}(P_1,\dots,P_s)$---quasi-polynomials introduced in~\cite{Norbury} that
count integer points in the interiors ${\mathcal M}_{g,s}$ of moduli spaces of Riemann surfaces of genus $g$ with $s>0$ holes with the
fixed perimeters $P_j\in{\mathbb Z}_+$, $j=1,\dots,s$ of holes in the Strebel uniformization.
Moreover, it was shown in~\cite{Ch95} that, when being expressed in the times $T^{\pm}_{2n}$
originated from the discretizations of the
moduli spaces, this model admits a decomposition into two Kontsevich models related by a Bogolyubov canonical
transformation---to the best of our knowledge it was the first example of the Givental-type
decomposition formulas~\cite{Giv}.

We use the decomposition formula of~\cite{Ch95} to express the terms ${\mathcal F}_{g,s}$
of the free energy expansion of this model in the
form of finite sums over graphs whose vertices are terms of the expansion of the free energy of Kontsevich
matrix model, internal edges correspond to quadratic terms in the canonical transformation operator, external
half edges (dilaton leaves) correspond to the constant shifts of the higher times, and external legs
(ordinary leaves) carry the times $T^{\pm}_{2n}$.
This graph representation provides another proof of quasi-polynomiality of $N_{g,s}(P_1,\dots,P_s)$
and that these quasi-polynomials depend only on even powers of $P_i$.

From \cite{Ey11} and \cite{DOSSIde} we know that the terms of topological recursion \cite{Ey},\cite{ChEy},\cite{CEO},\cite{AlMM} 
based on a certain
spectral curve satisfying a compatibility condition (relating the $w_{0,1}$ and $w_{0,2}$ invariants) describe ancestor invariants of a cohomological field theory, or equivalently a Frobenius manifold.

A fundamental family of Frobenius manifolds described by Dubrovin are Hurwitz spaces.  For $\mu=(\mu_1,\dots,\mu_n)$ define the Hurwitz space $H_{g,\mu}$  to consists of genus $g$ branched covers of the sphere with $n$ labeled points over $\infty$ of ramification profile $(\mu_1,\dots,\mu_n)$ and simple ramification over $\bp^1-\infty$.  It has dimension $|\mu|+n+2g-2$ where $|\mu|=\mu_1+\cdots+\mu_n$.

The 2-dimensional Hurwitz-Frobenius manifold $H_{0,(1,1)}$ consists of double branched covers of the sphere, with two branch points and unramified at infinity.  Its free energy is
\beq
F=\frac{1}{2}t_{0,0}^2t_{0,1}+\frac{1}{2}t_{0,1}^2\log{t_{0,1}}
\label{Frob}
\eeq
with the Euler vector field
$$E=t_{0,0}\frac{\partial}{\partial t_{0,0}}+2t_{0,1}\frac{\partial}{\partial t_{0,1}}.$$
Note that expression (\ref{Frob}) appears as a standard term (the perturbative part) in the
expansion of any matrix model upon identification of $t_{0,1}$ with the normalized number of eigenvalues.

We discuss the combinatorial relation between the \dvol\  and Gaussian means
using the cohomological field theory (CohFT) description, which relates
the \dvol\  to ancestor invariants of a CohFT. These ancestor invariants are evaluated already
in terms of the {\em closed} moduli spaces ${\mathcal M}_{g,s}$ compactified by Deligne and Mumford.
We thus provide explicit description of
these invariants in terms of the Gaussian means $W^{(g)}_{s}(x_1,\dots,x_s)$.

In~\cite{Ch93},~\cite{Ch95} the conjecture was put forward that the very same
KPMM expressed in times $T^{\pm}_{2n}$
describes the Deligne--Mumford stratification of the {\em closed} moduli spaces $\overline{\mathcal M}_{g,s}$ while
{\em reduction coefficients} $c_{g,s,r_q}$~\cite{Ch93} related to ancestor invariants in the above scheme
were conjectured to be positive rational numbers
proportional modulo some presribed combinatorial factors to the numbers of {\em screens}~\cite{McShane-Penner} for the corresponding Riemann surfaces. Although the original conjecture of~\cite{Ch93}, which assumed the existence of
covering tori for $\overline{\mathcal M}_{g,s}$, turned out to be incorrect,
this interpretation nevertheless helps to explain why the \dvol\  $N_{g,s}(P_1,\dots,P_s)$
are ${\mathbb Z}_2$ quasi-polynomials modulo the parity of $P_i$, not polynomials, and why the 
dependence is only on even powers of $P_i$: In formulas for the
moduli space multi-component reductions the two sets of times undergo symmetrization depending on the multicomponent
reduction structure, and  we therefore lack the single symmetrized expression when evaluating the inverse Laplace transform.

The paper is organized as follows. In Sec.~\ref{s:matrix-model} we establish the equivalence between the Gaussian
means (the correlation functions) and the terms of expansion of the KPMM free energy. 

We devote Sec.~\ref{s:KPMM} to describing results of~\cite{Ch93} and \cite{Ch95} concerning the open discrete
moduli spaces by which we produce the formula
relating the above Gaussian means and the \dvol\  in a purely combinatorial way. We also demonstrate how the
quantum curve can be obtained as a specialization of the KPMM to the case of unit size matrices. We describe the Givental-type
decomposition formulas for the KPMM obtained in \cite{Ch95} representing them in terms of graph expansions for
the free energy terms. We use this graph representation for proving the quasi-polynomiality of
the \dvol\  and for making a link to CohFT.  Finally, we discuss a link provided by this representation to
a Deligne--Mumford-type stratification of moduli spaces.

In Sec.~\ref{s:CohFT}, we identify the Gaussian means expansion terms with the ancestor invariants of
a cohomological field theory using the results of \cite{DNOPSSup} and \cite{DOSSIde}. The decomposition thus obtained has a canonical Givental form differing therefore with the decomposition in the preceding section. We show that the coefficients of 
this (``alternative'') decomposition, or equivalently, the coefficients of the
Laplace transforms of the quasi-polynomials $N_{g,s}(P_1,\dots,P_s)$,
are the special coefficients $b_k^{(g)}$, which are also the coefficients of $P_g(k)$ from \cite{APRW, ACRPS}, that represent in the ``most economic'' way to record the
genus filtered $s$-loop means $W_s^{(g)}(x_1,\dots, x_s)$---for example, for $s=1$, we need only $g$ coefficients 
$b_{k}^{(g)}$, $k=0, \ldots g-1$. We identify the coefficients $b_k^{(g)}$ with simple linear combinations of the CohFT ancestor invariants.

In Sec.~\ref{s:HZ}, we concentrate on the case of
a one-loop mean satisfying the Harer--Zagier (HZ) recurrence relation which is proven using integrals over $N\times N$ Hermitian matrices for fixed $N$.  On the basis of the original HZ relation,
we obtain a more effective recursion relation for the coefficients $b_k^{(g)}$, $k=0,\dots,g-1$,
of shapes and formulate the recursion procedure for obtaining the higher coefficients $b_{g-s}^{(g)}$ for a fixed
$s$ at all $g$. We prove that $b_k^{(g)}$, $k=0,\dots,g-1$, are positive integers for all $g$ and $k$, thus providing further evidence for the main conjecture of \cite{APRW}, which would imply this positivity property.
We present two alternative
derivations of the first subleading coefficient $b^{(g)}_{g-2}$: from the HZ
recurrence relations and from the decomposition formulas of Sec.\ref{s:KPMM}.


\section{The matrix-model representation for the multi-loop Gaussian means}
\label{s:matrix-model}

We consider a sum of connected diagrams with $s$ backbones, or loop insertions, each
carrying the corresponding variable $u_i$, $i=1,\dots,s$.
Our aim is to formulate the matrix model that describes all
genus-$g$ contributions in terms of {\em shapes} ---the connected fatgraphs of genus $g$ with $s$ faces and with vertices
of arbitrary order greater or equal three; according to the formula for the Euler characteristic, for a fixed $g$ and $s$ we then
have only a finite number of such graphs, and we let $\Gamma_{g,s}$ denote this finite set.  In fact, $\Gamma_{g,s}$
enumerates the collection of cells in the canonical (Strebel--Penner) ideal cell decomposition of moduli space ${\mathcal M}_{g,s}$,
the combinatorial part of which we shall recall here.

The origin of the word shapes here come from \cite{APRW}, where we considered the Poincare dual graphs of such fatgraphs, and by analogy with the arguments in that paper, one can easily see that $\Gamma_{g,s}$ is in bijection with circular chord diagrams which are also "shapes" in the terminology of \cite{APRW}, that is chord diagrams which are seeds and which has no one-chords, again in the terminology of \cite{APRW}.

\subsection{Multi-loop Gaussian means}
\label{ss:loop-means}

We begin by considering the correlation functions, or means, 
\be
\left\langle\prod_{i=1}^s(\tr H^{k_i})\right\rangle = \int_{H\in \mathcal{H}_{N}} \left(\prod_{i=1}^s\tr H^{k_i}\right) e^{-\frac N2 \tr H^2} DH,
\label{corr}
\ee
where $\mathcal{H}_{N}$ is the set of Hermitian $N\times N$ matrices. 
By Wick's theorem, any correlation function of the form (\ref{corr}) can be calculated as the sum over all possible (complete) pairings between matrix entries $M_{ij}$, and every pairing is given by just the two-point correlation function 
$\langle H_{i,j}H_{k,l}\rangle =\frac 1N \delta_{il}\delta_{jk}$. Such pairings gives rise to  \emph{fatgraphs} containing ordered set of $s$ vertices of valencies $k_i$, $i=1,\dots,s$, and $\sum_{i=1}^s k_i/2$ edges corresponding to the pairings (represented by double lines, thus inducing the fat graph structure). Furthermore for each vertex we also have a first incident edge given. We denote this set of fatgraphs $\widehat\Gamma(k_1,\dots,k_s)$. Then the above sum reads
$$
\left\langle\prod_{i=1}^s(\tr H^{k_i})\right\rangle = \sum_{\gamma\in \widehat\Gamma(k_1,\dots,k_s)}
N^{b(\gamma)-\sum_{i=1}^s k_i/2},
$$
where $b(\gamma)$ is the number of boundary components of $\gamma$.

Let $\widehat\Gamma(k_1,\dots,k_s)^c$ be the subset of $\widehat\Gamma(k_1,\dots,k_s)$ which consist of connected such fatgraphs and we define
$$\Bigl\langle\prod_{i=1}^s(\tr H^{k_i})\Bigr\rangle^{\mathrm{conn}} = \sum_{\gamma\in \widehat\Gamma(k_1,\dots,k_s)^c}
N^{b(\gamma)-\sum_{i=1}^s k_i/2}.$$
Then clearly
\be
\Bigl\langle\prod_{i=1}^s(\tr H^{k_i})\Bigr\rangle=\sum_{\small \mathrm{All}\ I_1\sqcup I_2\sqcup\cdots\sqcup I_p=\{1,\dots,s\}
\atop I_j\ne \emptyset}
\prod_{j=1}^p \Bigl\langle\prod_{i\in I_j}(\tr H^{k_i})\Bigr\rangle^{\mathrm{conn}}.
\label{total-vis-conn}
\ee
Second, we consider the $1/N$-expansion of the connected correlation function
\be
N^{s-2} \Bigl\langle\prod_{i=1}^s(\tr H^{k_i})\Bigr\rangle^{\mathrm{conn}}
=\sum_{g=0}^\infty N^{-2g} \Bigl\langle\prod_{i=1}^s(\tr H^{k_i})\Bigr\rangle_g^{\mathrm{conn}}
\ee
to segregate its part corresponding to connected fat graphs of genus $g$. Wick's theorem therefore implies that 
\be
\Bigl\langle\prod_{i=1}^s(\tr H^{k_i})\Bigr\rangle_g^{\mathrm{conn}} = 
 |\widehat\Gamma_g(k_1, \ldots, k_s)^c|, 
\label{sg2}
\ee
where $\widehat\Gamma_g(k_1, \ldots, k_s)^c$ is the subset of $\widehat\Gamma(k_1, \ldots, k_s)^c$ consisting of genus $g$ fatgraphs.
We now consider the union
$$ \widehat\Gamma_{g,s}^c = \bigsqcup_{\{k_1,\dots,k_s\}\in {\mathbb Z}_+^s} \widehat\Gamma_g(k_1, \ldots, k_s)^c.$$

\begin{figure}[h]
{\psset{unit=0.5}
\begin{pspicture}(-6.5,-5.5)(7.5,5.5)
\pscircle[linecolor=black,linewidth=6pt](0,0){4.15}
\pscircle[linecolor=white,linewidth=4pt](0,0){4.1}
\psbezier[linecolor=black,linewidth=6pt](-3.48,2)(-2,1.9)(2,1.9)(3.48,2)
\psbezier[linecolor=white,linewidth=4pt](-3.48,2)(-2,1.9)(2,1.9)(3.48,2)
\psbezier[linecolor=black,linewidth=6pt](-3.48,2)(-6.5,2)(-5,4.6)(-3.48,2)
\psbezier[linecolor=white,linewidth=4pt](-3.48,2)(-6.5,2)(-5,4.6)(-3.48,2)
\psbezier[linecolor=black,linewidth=6pt](0,-4)(0,-9)(-2.75,0.7)(-3.48,2)
\psbezier[linecolor=white,linewidth=4pt](0,-4)(0,-9)(-2.75,0.7)(-3.48,2)
\rput{150}(-3.48,2){\pcline[linewidth=1.5pt,linecolor=red](0,0)(1,0)}
\rput{150}(3.48,2){\pcline[linewidth=1.5pt,linecolor=red](0,0)(1,0)}
\rput{90}(0,-4){\pcline[linewidth=1.5pt,linecolor=red](0,0)(1,0)}
\rput(0,0){\pscircle[linecolor=black,fillstyle=solid,fillcolor=white](-3.48,2){0.4}}
\rput(0,0){\pscircle[linecolor=black,fillstyle=solid,fillcolor=white](3.48,2){0.4}}
\rput(0,0){\pscircle[linecolor=black,fillstyle=solid,fillcolor=white](0,-4){0.4}}
\rput(-3.48,2){\makebox(0,0)[cc]{\small 1}}
\rput(3.48,2){\makebox(0,0)[cc]{\small 3}}
\rput(0,-4){\makebox(0,0)[cc]{\small 2}}
\rput(4.5,0){\pcline[linewidth=3pt,linecolor=blue]{<->}(0,0)(2,0)}
\end{pspicture} }
{\psset{unit=0.5}
\begin{pspicture}(-5,-5.5)(5,5.5)
\psellipse[linewidth=6pt,linecolor=black](0,-1)(5.65,4.15)
\psellipse[linewidth=4pt,linecolor=white](0,-1)(5.6,4.1)
\psframe[linecolor=white, fillstyle=solid, fillcolor=white](-6,-1)(6,4)
\pcline[linewidth=1.5pt,linecolor=red](-3,5)(3,5)
\pcline[linewidth=1.5pt,linecolor=red](0,2)(3,2)
\pcline[linewidth=1.5pt,linecolor=red](-3,-1)(0,-1)
\psarc[linewidth=6pt,linecolor=black](-3,-1){.5}{180}{360}
\psarc[linewidth=4pt,linecolor=white](-3,-1){.5}{180}{365}
\psarc[linewidth=6pt,linecolor=black](-3,-1){1.5}{180}{360}
\psarc[linewidth=4pt,linecolor=white](-3,-1){1.5}{180}{365}
\psarc[linewidth=6pt,linecolor=black](0.5,-1){1}{180}{360}
\psarc[linewidth=4pt,linecolor=white](0.5,-1){1}{175}{360}
\pcline[linewidth=6pt,linecolor=black](1.5,2)(1.5,-1)
\pcline[linewidth=4pt,linecolor=white](1.5,2.05)(1.5,-1.05)
\psarc[linewidth=6pt,linecolor=black](0,2){.5}{180}{360}
\psarc[linewidth=4pt,linecolor=white](0,2){.5}{180}{365}
\psarc[linewidth=6pt,linecolor=black](3,2){.5}{180}{360}
\psarc[linewidth=4pt,linecolor=white](3,2){.5}{175}{360}
\psbezier[linecolor=black,linewidth=6pt](-3.5,-1)(-3.5,2)(-1.5,2)(-1.5,5)
\psbezier[linecolor=white,linewidth=4pt](-3.5,-1.05)(-3.5,2)(-1.5,2)(-1.5,5.05)
\psbezier[linecolor=black,linewidth=6pt](-4.5,-1)(-4.5,2)(-.5,2)(-.5,5)
\psbezier[linecolor=white,linewidth=4pt](-4.5,-1.05)(-4.5,2)(-.5,2)(-.5,5.05)
\psbezier[linecolor=black,linewidth=6pt](-5.5,-1)(-5.5,2)(-2.5,2)(-2.5,5)
\psbezier[linecolor=white,linewidth=4pt](-5.5,-1.05)(-5.5,2)(-2.5,2)(-2.5,5.05)
\psbezier[linecolor=black,linewidth=6pt](5.5,-1)(5.5,3)(2.5,3)(2.5,5)
\psbezier[linecolor=white,linewidth=4pt](5.5,-1.05)(5.5,3)(2.5,3)(2.5,5.05)
\psbezier[linecolor=black,linewidth=6pt](-.5,2)(-.5,3.5)(.5,3.5)(.5,5)
\psbezier[linecolor=white,linewidth=4pt](-.5,1.95)(-.5,3.5)(.5,3.5)(.5,5.05)
\psbezier[linecolor=black,linewidth=6pt](3.5,2)(3.5,3.5)(1.5,3.5)(1.5,5)
\psbezier[linecolor=white,linewidth=4pt](3.5,1.95)(3.5,3.5)(1.5,3.5)(1.5,5.05)
\rput(0,5.6){\pscircle[linecolor=black,fillstyle=solid,fillcolor=white](0,0){0.4}}
\rput(-1.5,-0.4){\pscircle[linecolor=black,fillstyle=solid,fillcolor=white](0,0){0.4}}
\rput(1.5,2.6){\pscircle[linecolor=black,fillstyle=solid,fillcolor=white](0,0){0.4}}
\rput(0,5.6){\makebox(0,0)[cc]{\small 1}}
\rput(1.5,2.6){\makebox(0,0)[cc]{\small 3}}
\rput(-1.5,-0.4){\makebox(0,0)[cc]{\small 2}}
\end{pspicture} }

\caption{\small Transition from ciliated multivalent vertices to linear backbones.}
\label{fi:backbones}
\end{figure}
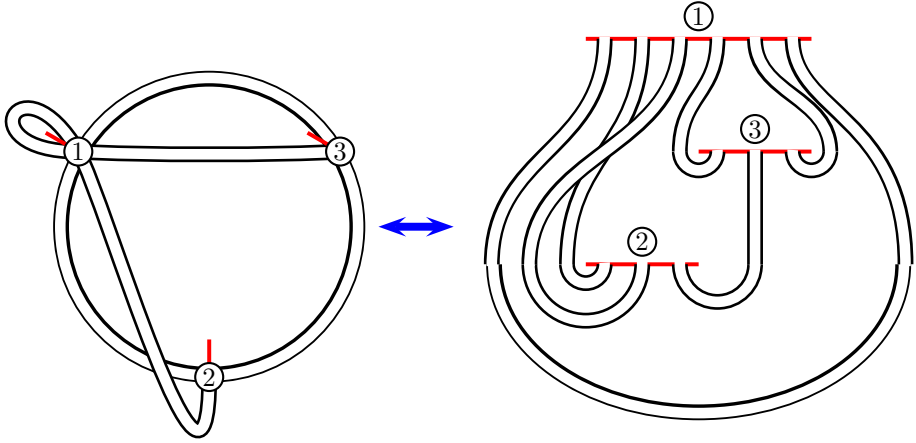

By blowing up each vertex to a backbone we exactly get the genus $g$ chord diagram on $s$ linear backbones as considered in \cite{APRW}, \cite{ACRPS}
(see Fig.~\ref{fi:backbones}).

We have the following formula
\bea
(-1)^s\left\langle\prod_{i=1}^s\tr\log(1-u_iH)\right\rangle_g^{\mathrm{conn}}&=&\sum_{\{k_1,\dots,k_s\}\in {\mathbb Z}_+^s}
\prod_{i=1}^s\Bigl(\frac{u_i^{k_i}}{k_i}\Bigr)\left\langle\prod_{i=1}^s(\tr H^{k_i})\right\rangle_g^{\mathrm{conn}}\nonumber\\
&=&\sum_{\{k_1,\dots,k_s\}\in {\mathbb Z}_+^s} \sum_{\gamma \in \Gamma_g(k_1,\dots,k_s)^c} \frac {1}{|\hbox{Aut\,}(\gamma) |}
\prod_{i=1}^s u_i^{k_i},
\label{det-rep}
\eea
where $ \Gamma_g(k_1,\dots,k_s)^c$ is the set of connected fat graphs of genus $g$ with $s$ (nonciliated) ordered vertices of 
valencies $k_1,\ldots,k_s$ and $\hbox{Aut\,}(\gamma)$ is the automorphism groups of the fatgraph $\gamma$ which respects the ordering of the vertices.
We obtain the actual multi-loop case (up to overall sign) by differentiation:
\beq
\label{loop-rep}
\left\langle\prod_{i=1}^s\tr\frac{1}{I-u_iH}\right\rangle_g^{\mathrm{conn}}\equiv
\left\langle\prod_{i=1}^s\tr\biggl[\sum_{k_i=1}^\infty u_i^{k_i}H^{k_i}\biggr]\right\rangle_g^{\mathrm{conn}}=
(-1)^s\biggl[\prod_{i=1}^s u_i\frac{\partial}{\partial u_i}\biggr] \left\langle\prod_{i=1}^s\tr\log(1-u_iH)\right\rangle_g^{\mathrm{conn}}.
\eeq
 By combining formula (\ref{sg2}) with (\ref{loop-rep}), we find that 
 
 $$\biggl[\prod_{i=1}^s u_i\frac{\partial}{\partial u_i}\biggr] \left\langle\prod_{i=1}^s\tr\log(1-u_iH)\right\rangle_g^{\mathrm{conn}} = \sum_{\gamma\in \widehat\Gamma_{g,s}^c} N^{2-2g}\prod_{i=1}^s u_i^{k_i}.$$

In order to proceed to the
Poincar\'e dual fatgraphs, which provides an equivalent description of chord diagrams in terms of \emph{shapes} understood as closed fatgraphs with vertices of order not lower than three, it is more instructive to consider the non-ciliated fat graphs and perform the proper resummation (see the next subsection).
We observe that the effect of differentiating w.r.t. the variables $u_i$ as in (\ref{loop-rep}) is precisely to add a cilium at vertex $i$.


\subsection{Summing up rainbow and ladder diagrams---formulating the matrix model}
\label{ss:summinf-up}

\begin{figure}[h]
{\psset{unit=1}
\begin{pspicture}(-7,-1)(7,1)
\pcline[linewidth=1pt](-6,0)(-5,0)
\rput(-5.5,-0.7){\makebox(0,0){$1$}}
\rput(-4.5,0){\makebox(0,0){$+$}}
\pcline[linewidth=1pt](-4,0)(-2.5,0)
\rput(-3.25,-0.7){\makebox(0,0){$u^2$}}
\rput(-2,0){\makebox(0,0){$+$}}
\psarc[linewidth=1.5pt,linestyle=dashed](-3.25,0){.4}{0}{180}
\pcline[linewidth=1pt](-1.5,0)(0.2,0)
\rput(-.65,-0.7){\makebox(0,0){$u^4$}}
\rput(0.7,0){\makebox(0,0){$+$}}
\psarc[linewidth=1.5pt,linestyle=dashed](-1,0){.3}{0}{180}
\psarc[linewidth=1.5pt,linestyle=dashed](-.3,0){.3}{0}{180}
\pcline[linewidth=1pt](1.2,0)(2.7,0)
\rput(1.95,-0.7){\makebox(0,0){$u^4$}}
\rput(3.2,0){\makebox(0,0){$+$}}
\psarc[linewidth=1.5pt,linestyle=dashed](1.95,0){.3}{0}{180}
\psarc[linewidth=1.5pt,linestyle=dashed](1.95,0){.5}{0}{180}
\pcline[linewidth=1pt](3.5,0)(5.6,0)
\rput(4.5,-0.7){\makebox(0,0){$u^6$}}
\rput(6.5,0){\makebox(0,0){$+\dots\equiv$}}
\psarc[linewidth=1.5pt,linestyle=dashed](4.15,0){.3}{0}{180}
\psarc[linewidth=1.5pt,linestyle=dashed](4.15,0){.45}{0}{180}
\psarc[linewidth=1.5pt,linestyle=dashed](5.05,0){.35}{0}{180}
%
\psframe[linewidth=1pt](7.5,-0.1)(8.6,0.1)
\pcline[linewidth=1pt](7.4,-0.1)(8.7,-0.1)
\pcline[linewidth=1pt](7.4,0.1)(8.7,0.1)
\rput(8,-0.7){\makebox(0,0){$f(u)$}}
\end{pspicture} }
\caption{\small The procedure of summing up rainbow diagrams of chords (dashed lines) for a
single backbone (solid lines). The result is the new edge of the backbone
indicated by a double line (the thickened edge).}
\label{fi:rainbow}
\end{figure}
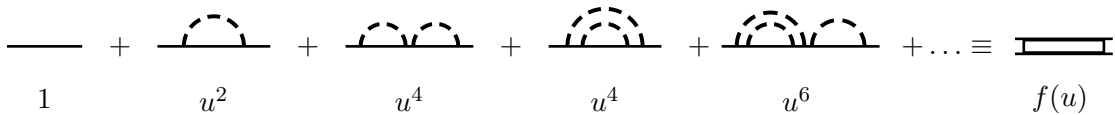

In the expression (\ref{det-rep}), we may perform a resummation
over planar diagrams.  To this end, we shall refer to a planar  chord diagram on an interval
as a {\it rainbow diagram} as depicted in Fig.~\ref{fi:rainbow}.
The number of rainbow diagrams 
of course have the generating
function
\beq
\label{factor-f}
f(u_i):=\frac{1-\sqrt{1-4u_i^2}}{2u_i^2},
\eeq
so summing over rainbow diagrams is thus mere Catalan number counting, which effectively reduces to
replacing the original edge of a chord diagram by a thickened edge carrying
the factor $f(u_i)$.

\begin{figure}[h]
{\psset{unit=0.5}
\begin{pspicture}(-10,-5)(4,2.5)
\newcommand{\PATTERN}{%
\psframe[linewidth=1pt,linecolor=white,fillstyle=crosshatch,hatchwidth=0.5pt](0.1,-0.1)(4,-4)
\pswedge[linecolor=white,fillstyle=solid,fillcolor=white](0,-5){2.2}{0}{90}
\pswedge[linecolor=white,fillstyle=solid,fillcolor=white](5,0){2.2}{180}{270}
\pswedge[linecolor=white,fillstyle=solid,fillcolor=white](0,0){2.2}{270}{360}
\pswedge[linecolor=white,fillstyle=solid,fillcolor=white](3.5,-3.5){1.2}{30}{250}
\pcline[linewidth=1.5pt,linestyle=dashed](5,0)(3.5,-3.5)
\pcline[linewidth=1.5pt,linestyle=dashed](0,-5)(3.5,-3.5)
\pscircle[linecolor=white,fillstyle=solid,fillcolor=white](3.5,-3.5){.7}
\pswedge[linecolor=white,fillstyle=solid,fillcolor=white](0,-5){1.7}{0}{90}
\pswedge[linecolor=white,fillstyle=solid,fillcolor=white](5,0){1.7}{180}{270}
\pscircle(0,0){2}
\pscircle(0,0){1.7}
\psarc[linewidth=1pt]{->}(0,0){1.3}{250}{290}
\psarc[linewidth=1pt](0,-5){2}{10}{170}
\psarc[linewidth=1pt](0,-5){1.7}{10}{170}
\psarc[linewidth=1pt]{->}(0,-5){1.3}{70}{110}
\psarc[linewidth=1pt](-4.33,-2.5){2}{-45}{45}
\psarc[linewidth=1pt](-4.33,-2.5){1.7}{-45}{45}
\psarc[linewidth=1pt]{->}(-4.33,-2.5){1.3}{-20}{20}
\psarc[linewidth=1pt](5,0){2}{135}{260}
\psarc[linewidth=1pt](5,0){1.7}{135}{260}
\psarc[linewidth=1pt]{->}(5,0){1.3}{160}{200}
\psarc[linewidth=1pt](3.5,-3.5){1}{30}{250}
\psarc[linewidth=1pt](3.5,-3.5){.7}{30}{250}
\psarc[linewidth=1pt]{->}(3.5,-3.5){.4}{90}{200}
\pcline[linewidth=1.5pt,linestyle=dashed](1.7,0)(3.3,0)
\pcline[linewidth=1.5pt,linestyle=dashed](-1.47,-0.85)(-2.85,-1.65)
\pcline[linewidth=1.5pt,linestyle=dashed](-1.47,-4.15)(-2.85,-3.35)
\rput{60}(0,0){\pcline[linewidth=1.5pt,linestyle=dashed](1.7,0)(2.5,0)}
\rput{110}(0,0){\pcline[linewidth=1.5pt,linestyle=dashed](1.7,0)(2.5,0)}
\rput{165}(0,0){\pcline[linewidth=1.5pt,linestyle=dashed](1.7,0)(2.5,0)}
\rput(0,0){\makebox(0,0)[cc]{$i$}}
\rput(0,-4.5){\makebox(0,0)[cc]{$j$}}
\rput(4.5,0){\makebox(0,0)[cc]{$l$}}
\rput(3.7,-3.7){\makebox(0,0)[cc]{$m$}}
}
\rput(-5,0){\PATTERN}
\rput(5,0){\PATTERN}
\rput(1,-1){\makebox(0,0)[cc]{$+$}}
\rput(11.5,-1){\makebox(0,0)[cc]{$+\cdots=$}}
\rput(-5,0){\pcline[linewidth=1.5pt,linestyle=dashed](0,-1.7)(0,-3.3)}
\rput(-5,0){\psarc[linewidth=3pt,linecolor=red](0,0){1.85}{210}{270}}
\rput(-5,0){\psarc[linewidth=3pt,linecolor=red](0,-5){1.85}{30}{90}}
\rput(5,0){\psarc[linewidth=3pt,linecolor=red](0,0){1.85}{210}{280}}
\rput(5,0){\psarc[linewidth=3pt,linecolor=red](0,-5){1.85}{30}{100}}
\rput(5,0){\psframe[linewidth=1pt,linecolor=white,fillstyle=solid,fillcolor=white](0.05,-2.1)(0.6,-2.9)}
\rput(5,0){\psarc[linewidth=1.5pt,linestyle=dashed](2,-2.5){2.5}{160}{200}}
\rput(5,0){\psarc[linewidth=1.5pt,linestyle=dashed](-2,-2.5){2.5}{-20}{20}}
\rput(17,0){
\pcline[linewidth=10pt,linecolor=red](-2,-2)(2,-2)
\pcline[linewidth=10pt,linecolor=blue,linestyle=dashed](2,-2)(2.7,1)
\pcline[linewidth=10pt,linecolor=blue,linestyle=dashed](2,-2)(2,-5)
\pcline[linewidth=10pt,linecolor=blue,linestyle=dashed](2,-2)(5,-2.5)
\pcline[linewidth=10pt,linecolor=blue,linestyle=dashed](-2,-2)(-3.5,1)
\pcline[linewidth=10pt,linecolor=blue,linestyle=dashed](-2,-2)(-3.5,-5)
\pcline[linewidth=7pt,linecolor=white](-2,-2)(2,-2)
\pcline[linewidth=7pt,linecolor=white](2,-2)(2.7,1)
\pcline[linewidth=7pt,linecolor=white](2,-2)(2,-5)
\pcline[linewidth=7pt,linecolor=white](2,-2)(5,-2.5)
\pcline[linewidth=7pt,linecolor=white](-2,-2)(-3.5,1)
\pcline[linewidth=7pt,linecolor=white](-2,-2)(-3.5,-5)
\rput(0,0){\makebox(0,0)[cc]{$i$}}
\rput(0,-4.5){\makebox(0,0)[cc]{$j$}}
\rput(4.5,0){\makebox(0,0)[cc]{$l$}}
\rput(3.7,-3.7){\makebox(0,0)[cc]{$m$}}
}
\end{pspicture} }
\caption{\small The procedure of performing a sum over ladder diagrams. The thickened edges
associated with the selected ladder are painted dark. The crosshatched domain will become a four-valent vertex in the
new fat graph, the white domain to the left of the crosshatched one becomes a three-valent vertex, etc. Ladders turn into
edges of the new fat graph. The double solid line in the right-hand side of the figure is
the edge obtained from darkened thickened edges in the left-hand side of the figure.}
\label{fi:cycles}
\end{figure}
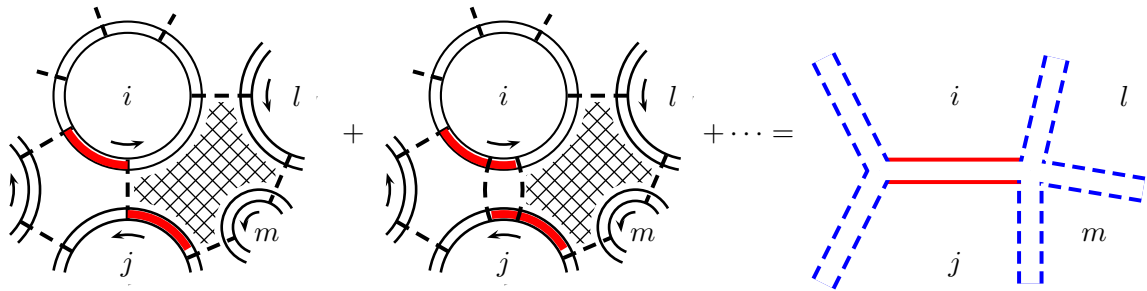

We now consider the summation of ladder-type diagrams, where
a ``rung'' of the ladder joins two cycles that carry in general distinct (but maybe coincident) indices
$i$ and $j$ (see an example in Fig.~\ref{fi:cycles}).
Each ladder contains at least one rung, which is
a chord carrying the factor $u_iu_j$. 
We obtain the set of {\em new edges} and
{\em new vertices} if we further blow up cycles of double-line backbone edges
until they will be joined pairwise along rungs (each containing at least one rung);
disjoint parts of these cycles will then constitute loops of lengths $2r_k\ge 6$ alternatively bounded by $r_k$ rungs
(the chords) and
$r_k$ double lines (the thickened edges of circular backbones,
possibly with different backbone indices of different edges);
these loops then become vertices of the respective orders $r_k\ge 3$ of the {\em new fat graph}. 

The diagrammatic expression (Fig.~\ref{fi:cycles}) demonstrates that we have exactly one factor $f(u_i)$ and
one factor $f(u_j)$ per each rung of a ladder (we introduce the same cyclic orientations on all circular backbones):
these are the factors associated with those thickened edges of backbones that precede the points of attachments of
the rung to the $i$th and $j$th backbones. This assignment is thus uniquely determined and no thickened
propagators remain unassigned, so the corresponding sum of ladder diagrams reads


\beq
\label{KP-rung}
\sum_{k=1}^\infty (u_iu_jf(u_i)f(u_j))^k=\frac{1}{(u_if(u_i)u_jf(u_j))^{-1}-1}:=\frac{1}{e^{\lambda_i+\lambda_j}-1},
\eeq
where we have introduced
\beq
\label{u-2-lambda}
e^{\lambda_i}=\frac{1+\sqrt{1-4u_i^2}}{2u_i},\quad\hbox{or}\quad u_i=\frac{1}{e^{\lambda_i}+e^{-\lambda_i}}.
\eeq

We arrive at the main result of this section: 

\begin{theorem}\label{lm:diagrams}
The genus-$g$ term of the (nonciliated) $s$-backbone case is given by the following
(finite!) sum over fatgraph shapes $\Gamma_{g,s}$ of genus $g$ with $s$ edges whose
vertices have valences at least three:
\beq
\Bigl\langle\prod_{i=1}^s\tr\log(e^{\lambda_i}+e^{-\lambda_i}-H)\Bigr\rangle_g^{\mathrm{conn}}
=\sum_{{\mathrm{all\ fatgraphs} }\atop \gamma~\in~\Gamma_{g,s}}
\frac{1}{| \Aut (\gamma)|} \prod_{\mathrm{edges}}\frac{1}{\e^{\lambda_e^{(+)}+\lambda_e^{(-)}}-1}
:= F^{(g)}_s(\lambda_1,\dots,\lambda_s),
\label{new-model}
\eeq
where $\pm$ denotes the two sides (faces) of the edge $e$.
The quantity $F^{(g)}_s(\lambda_1,\dots,\lambda_s)$ in the right-hand side is the term in the
diagrammatic expansion of the Kontsevich--Penner matrix model~\cite{ChMak1} given by the following normalised integral
over Hermitian $N\times N$-matrices $X$:
\beq
{\mathcal Z}[\Lambda]:=e^{\sum_{g,s}N^{2-2g}(\alpha/2)^{2-2g-s}F^{(g)}_s(\lambda)}=
\frac{\int DX e^{-\alpha N\tr\bigl[\frac14 \Lambda X\Lambda X+\frac12\log(1-X)+X/2 \bigr]}}
{\int DX e^{-\alpha N\tr\bigl[\frac14 \Lambda X\Lambda X-\frac14 X^2\bigr]}},
\label{KPMM-1}
\eeq
where the sum ranges over all stable curves $(2g+s>2)$ and $\Lambda$ is a diagonal matrix with
the numbers $e^{\lambda_i}$ on the diagonal.
\end{theorem}


The {\bf proof}  beyond the earlier works \cite{ChMak1}, \cite{PenPer} has essentially been given already.  
Let us further simply observe that in replacing boundary cycles in ladder diagrams by punctures and rungs by
arcs connecting those punctures, we reproduce the small elaboration of the original arc families in
\cite{Penner} where parallel copies of arcs are permitted in the current instance.  The underlying idea is then completely
obvious graphically, where (possibly empty) copies of boundary-parallel rainbow diagrams are here permitted in between arcs.
Notice that the old dual fatgraph
to the arc family in the original sense of \cite{PenPer} is the new fatgraph of the field theory defined above.

Differentiating the relation (\ref{new-model}) w.r.t.\ $\lambda_i$ on the right-hand side we obtain the standard
loop means, or (connected) correlation functions $W_s^{(g)}(x_1,\dots,x_s)$, $x_i=e^{\lambda_i}+e^{-\lambda_i}$,
of the Gaussian matrix model enjoying the standard recursion relations of the
topological expansion~\cite{Ey},~\cite{ChEy}. As an immediate consequence of the previous result, we have

We introduce the standard definition of the connected correlation functions (the $s$-loop means, or resolvents) of
the Gaussian matrix model with the potential $\frac{N}2\tr H^2$:
\bea
W_s^{(g)}(x_1,\dots,x_s)&:=&N^{s-2}\left\langle\prod_{i=1}^s\tr\frac{1}{x_i-H}\right\rangle_g^{\mathrm{conn}}\nonumber\\
&=&N^{s-2}\sum_{\{k_1,\dots,k_s\}\in {\mathbb Z}_+^s}
\prod_{i=1}^s x_i^{-k_i-1}\left\langle\prod_{i=1}^s(\tr H^{k_i})\right\rangle_g^{\mathrm{conn}}
+x_1\delta_{s,1}\delta_{g,0}.
\label{resolvent}
\eea
Although the difference between thus defined quantities and original means (\ref{det-rep}) as well as with the generating functions $C_{g,s}(z)$
for $s$-backbone chord diagrams of \cite{APRW} and \cite{ACRPS} is only in changing the normalisation (see Remark~\ref{rm:Pg1})
precisely these resolvents (\ref{resolvent}) turn into symmetrical $s$-forms in the formalism of topological recursion of \cite{Ey}, \cite{ChEy} thus playing an instrumental role in this formalism.

\begin{corollary}\label{cor:loopmean}
The exact relation between resolvents (\ref{resolvent}) and the terms
of the free-energy expansion of the Kontsevich--Penner matrix model reads
\beq
W_s^{(g)}(e^{\lambda_1}+e^{-\lambda_1},\dots,e^{\lambda_s}+e^{-\lambda_s})=
\prod_{i=1}^s\left[\frac{1}{e^{\lambda_i}-e^{-\lambda_i}}\frac{\pa}{\pa \lambda_i}\right]F^{(g)}_s(\lambda_1,\dots,\lambda_s).
\label{loopmeans-KPMM}
\eeq
The quantities $W_s^{(g)}(x_1,\dots,x_s)$ here
enjoy the standard topological recursion~\cite{ChEy}, \cite{AlMM} for the spectral curve $x=e^{\lambda}+e^{-\lambda}$,
$y=\frac12 \bigl(e^{\lambda}-e^{-\lambda}\bigr)$.
\end{corollary}

\section{Matrix models and geometry of moduli spaces}
\label{s:KPMM}

\subsection{The Kontsevich construction for evaluating intersection indices}\label{ss:Kontsevich}

We begin with recalling the cell decomposition of moduli spaces of Riemann surfaces of genus $g$ with
$s>0$ marked points which was proved independently by Harer \cite{HarVir}, who presented a proof of Mumford using Strebel differentials~\cite{Strebel}, and by Penner \cite{Penner} using hyperbolic geometry.  This cell decomposition theorem
states that strata in the cell decomposition of the direct
product ${\mathcal M}_{g,s}\times {\mathbb R}^s_{+}$ of the open moduli space and the $s$-dimensional space of
strictly positive perimeters of holes
are in one-to-one correspondence with fat graphs of genus $g$ with $s$ faces (those are the shapes from
Sec.~\ref{s:matrix-model}) whose edges are decorated with strictly positive numbers $l_i\in {\mathbb R}_{+}$.
The perimeters $P_I$, $I=1,\dots,s$ are the sums of $l_i$ taken (with multiplicities) over edges incident to the
corresponding face (boundary component, or hole).  So it is natural to call them the {\em lengths} of the corresponding edges.

One associates the Chern classes to the corresponding holes, and Kontsevich~\cite{Kon88} found a remarkably simple
formula for representatives of these Chern classes $\psi_I$:
\beq
\psi_I=\sum_{k<j}^{n_I}d\left[\frac{l_{i_k}}{P_I}\right]\wedge d\left[\frac{l_{i_j}}{P_I}\right].
\label{omega}
\eeq
where $P_I=\sum_{j=1}^{n_I} l_{i_j}$ and  $i_1,\ldots,i_{n_l}$ 
enumerates edges incident to the $I$'th boundary component, which is therefore an $n_I$-gone and these edges are
linearly ordered in accordance with the orientation. Changing the starting edge results in adding an exact 2-form to (\ref{omega}) thus not affecting the result of integration. The quantities to be calculate are the
{\em intersection indices}
$$
\left\langle\tau_{d_1}\cdots\tau_{d_s}\right\rangle_g;=\int_{\overline{\mathcal M}_{g,s}}\prod_{I=1}^s \psi_I^{d_I},
$$
which do not depend on actual values of $P_I$ being purely cohomological objects.
The useful tools that enables the determination of the generating function for these numbers are, first, to multiply every $\psi_I^{d_I}$ by
$P_I^{2d_I}$ and, second, to perform the {\em Laplace transformation} w.r.t. all $P_I$, that is, to integrate over
the total direct product $\overline{\mathcal M}_{g,s}\times {\mathbb R}^s_{+}$. All unwanted factors arising when
the external derivative $d$ acts on $P_I$ in the denominators are then canceled by the integration measure
$dP_1\wedge \cdots \wedge dP_s$, and only the total volume form coming
with the factor $2^{2g-2+s}$  remains in the integral, which therefore
transforms into the diagrammatic expansion of the Kontsevich matrix model; this procedure was extensively
reviewed in the literature (see, e.g.,~\cite{LandoZvonkin}). On the other hand, this integration merely gives
\beq
\int_0^\infty\cdots\int_0^\infty dP_1\cdots dP_s e^{-\sum_I P_I\lambda_I}
\int_{\overline{\mathcal M}_{g,s}}\prod_{I=1}^s P_I^{2d_I}\psi_I^{d_I}=
\left\langle\tau_{d_1}\cdots\tau_{d_s}\right\rangle_g \prod_{I=1}^s \prod_{I=1}^s \frac{(2d_I)!}{\lambda_I^{2d_I+1}}.
\label{Kontsevich}
\eeq

The left-hand side of (\ref{Kontsevich}) can be presented as the sum over three-valent fat graphs with the
weights $1/(\lambda_{I_1}+\lambda_{I_2})$ on edges where $I_1$ and $I_2$ are indices of two (possibly coinciding)
cycles incident to a given edge. Also a factor $2^{|L|-|V|}$ appears (where $|V|$ and $|L|$ are the cardinalities
of the respective sets of vertices and edges).
The generating function is then the celebrated {\em Kontsevich matrix model}
\beq
e^{{\mathcal F}_{\text{K}}(\{\xi_k\})}:=\frac{\int DX e^{-\alpha N\tr\bigl[\frac12 X^2\Lambda +X^3/6 \bigr]}}
{\int DX e^{-\alpha N\tr\bigl[\frac12 X^2\Lambda\bigr]}},
\label{Kont}
\eeq
where
\beq
\xi_k:=\frac1N\sum_{i=1}^N\frac{(2k)!}{\lambda_i^{2k+1}}
\label{times-K}
\eeq
are the {\em times} of the Kontsevich matrix model.

\subsection{The discretization of moduli spaces}

The material in this subsection (and only in this subsection) is not rigorous and we need it only as a motivation for what to follow.
We suppose here the existence of a discretised version of the Kontsevich construction in
Subsection~\ref{ss:Kontsevich} and that such a hypothetical theory would also make sense on the boundary.

As was proposed in~\cite{Ch93}, set all the lengths of edges of the Penner--Strebel graphs to be nonnegative {\em integers}
$l_i\in {\mathbb Z}_+$, $i=1,\dots, |L|\le 6g-6+3s$.

Were a proper cohomological theory of discretely acting operators to exist, we could
introduce special representatives of the corresponding Chern classes $\widehat\psi_I$ in the spirit of
the original Kontsevich construction~\cite{Kon88} introducing also the analogue of the $\Omega$ two-form
\beq
\widehat\Omega =\sum_I P_I^2 \widehat\psi_I.
\label{hatOmega}
\eeq


We would then
replace integrations w.r.t. $l_i$ by half-infinite sums over integer points; in \cite{Ch93}
we called these lattices endowed with discrete integration the {\em discrete
moduli spaces} ${\mathcal M}^{\disc}_{g,s}$.

When integrating the above forms over the discrete moduli spaces, as in the case of the standard moduli spaces,
the integral
$\int_{{\mathcal M}^{\disc}_{g,s}}\prod_I P_I^{2d_I}\hat\psi_I^{d_I}$ does not make sense until we (i) include the closure of the moduli space and (ii) perform the discrete Laplace transformation w.r.t. the perimeters $P_I$
weighted with the standard function $e^{-\sum_I P_I\lambda_I}$ thus reconstructing, as in the original Kontsevich
approach, the total volume form on the totally discretized space
$\overline{\mathcal M}^{\disc}_{g,s}\times {\mathbb Z}_+^{s}$. Then, as in the Kontsevich case, we have
the purely combinatorial relation between the highest-order forms,
\beq
dP_1\wedge \cdots\wedge dP_s\wedge \prod_{I=1}^s P_I^{2d_I}\hat\psi_I^{d_I}
=\pm dl_1\wedge\cdots \wedge dl_{6g-6+3s}2^{2g-2-s},
\label{rel-forms}
\eeq
which does not depend on the numbers $d_I$ (provided $\sum_I d_I=3g-3+s$) and on the combinatorial type of the fat graph on
whose edges have length $l_i$.


An important subtlety is that since the length $l_i$ of every edge enters twice in the sum $\sum_{I=1}^s P_I$,
this sum is always a positive even number, and we must take this restriction into account
when performing the discrete Laplace transformation.

Because the {\em Deligne--Mumford compactification} of the moduli space is purely combinatorial, properly defined integrals
over continuous and discrete moduli spaces must match each other, and instead of the formula (\ref{Kontsevich})
we presumably obtain a formula of the type
\bea
&{}&
\sum^\infty_{{P_I=1 \atop \sum P_I\in 2{\mathbb Z}_{+}}} \prod_{I=1}^s P_I^{2d_I} e^{\sum_I P_I\lambda_I}
\int_{\overline{\mathcal M}^{\disc}_{g,s}}\prod_{I=1}^s \hat\psi_I^{d_I}\nonumber\\
&{}&\quad =\frac12 \sum_{P_I\in {\mathbb Z}_{+}}\prod_{I=1}^s P_I^{2d_I} e^{\sum_I P_I\lambda_I}(1+(-1)^{\sum_I P_I})
\left\langle\tau_{d_1}\cdots\tau_{d_s}\right\rangle_g \nonumber\\
&{}&\quad
=\frac12 \left\langle\tau_{d_1}\cdots\tau_{d_s}\right\rangle_g
\left[\prod_{I=1}^s T^-_{2d_I}(\lambda_I)+ \prod_{I=1}^s T^+_{2d_I}(\lambda_I)\right],
\nonumber
\eea
where the new times
\beq
T^\pm_{2k}(\lambda_I):=\frac{\pa^{2k}}{\pa \lambda^{2k}_I}\frac{1}{\mp e^{\lambda_I} -1}=\sum_{P_I=1}^\infty (\mp 1)^{P_I} P_I^{2k} 
e^{-\lambda_I P_I}
\label{times-KP}
\eeq
are discrete Laplace transforms.

In the next subsection, we demonstrate that the KPMM (\ref{KPMM-1}) corresponds to integrations over the {\em open} discrete moduli spaces.
Note here that 
following the Deligne and Mumford approach, we can represent integration over the open moduli spaces ${\mathcal M}_{g,s}$
(discrete or continuous) using the stratification of closed moduli spaces, as an alternating sum over reduction components,
\beq
\int_{{\mathcal M}_{g,s}}=\int_{{\overline{\mathcal M}}_{g,s}}+\sum_{q{\text{-component}}\atop {\text{reductions}} r}
(-1)^{|r|}c^{g,s}_r \prod_{l=1}^q \int_{{\overline{\mathcal M}}_{g_l,s_l}},
\label{strata}
\eeq
where $|r|$ is the reduction degree equal to $3g-3+s-\sum_l (3g_l-3+s_l)$ and $c^{g,s}_r$ are positive rational numbers
counting, roughly speaking, the numbers of the given type of reduction (boundary strata) per copy of 
${\overline{\mathcal M}}_{g,s}$ in the corresponding cell decompsition of ${\overline{\mathcal M}}_{g,s}$ (also related to the
number of {\em screens} \cite{McShane-Penner}).

Would the Kontsevich symplectic structure have an analytic continuation to the boundary, we might expect that
we can use  (\ref{strata}) in order to present an ``integration'' of the form $e^{\widehat\Omega}$ 
(see (\ref{hatOmega})) over an open moduli space ${\mathcal M}_{g,s}$ in the form of
alternative sums of products of integrations over closed moduli spaces  thus obtaining that
\bea
&{}&\int_{{\mathcal M}^{\disc}_{g,s}}e^{\widehat\Omega} 
=\left\langle \tau_{d_1}\cdots \tau_{d_s}\right\rangle_g \Bigl[\prod_{i=1}^s T^+_{2d_i} +\prod_{i=1}^s T^-_{2d_i}\Bigr]
\nonumber\\
&{}&+\sum_{q{\text{-component}}\atop {\text{reductions}} r} (-1)^{|r|}c^{g,s}_r \prod_{l=1}^q\biggl[
\left\langle \tau_{d_1}\cdots \tau_{d_{s_l}}\right\rangle_{g_l} \Bigl[ \prod_{k=1}^j T^+_{2d_k} \prod_{k=j+1}^{s_l} D^+_{2d_k} 
+ \prod_{k=1}^j T^-_{2d_k} \prod_{k=j+1}^{s_l} D^-_{2d_k}\Bigr] 
\biggr]
\label{conj}
\eea


where $\left\langle \tau_{d_1}\cdots \tau_{d_l}\right\rangle_g$ are the standard Kontsevich intersection indices and we 
segregate $\psi$-classes on reduction components into two groups: the original $\psi$-classes (their total number is always $s$) and those obtained via reductions (we put dilaton leaves $D^\pm_{2d_k}$ in correspondence with these classes; originally all
those classes were assumed to be just the trivial classes $\tau_0$). Note that the symmetrisation w.r.t. $T^+$ and $T^-$ occurs in every reduction component separately, so the result contains mixed terms of $T^+$ and $T^-$ thus being a quasi-polynomial.


Although the naive form of (\ref{conj}) does not hold, notably, a factorisation of $T^\pm$ into expressions 
independently symmetrized in each reduction component fails and we have to introduce relations between dilaton leaves at different 
components of the reduction, thus producing (inner) edges of a special graph representation in Subsec.~\ref{ss:discretemoduli},
the genuine resulting expression in terms of graphs presented in this subsection manifests at least some of the properties
we might expect from (\ref{conj}): it depends only on even times $T^\pm_{2k}(\lambda)$ and it exhibits an alternating sign structure w.r.t. the total powers of monomials in these times (see Remark \ref{rm:alternative}), so we expect that this model contains some 
useful information about the cell decomposition of the closed moduli spaces, and not only that of the open moduli spaces, on which we
concentrate in what follows.


\subsection{Open discrete moduli spaces and the Kontsevich--Penner matrix model}

No matrix-model description of integrations over the closure of the moduli spaces exists
(and  would presumably just be
obtain from the original Kontsevich intersection indices weighted by different times). 
On the other hand, integrals over discrete open moduli spaces 
are well defined for {\em volumes} of discrete
open moduli spaces ${\mathcal M}^{\disc}_{g,s}$.


Following \cite{Norbury} we thus define the {\em \dvol\ } 
$N_{g,s}(P_1,\dots,P_s)$ which is a weighted count of the integer points inside
${\mathcal M}^{\disc}_{g,s}\times {\mathbb Z}_+^{s}$ for fixed positive integers $P_I$, $I=1,\dots,s$, which are
the perimeters of the holes (cycles). These numbers are equal (modulo the standard factors of volumes of automorphism
groups) to the numbers of all fat graphs with vertices of valencies three and higher and with positive integer
lengths of edges subject to the restriction that the lengths of all cycles are fixed. Using the above identity
$$
\sum_{I=1}^s \lambda_I P_I=\sum_{e\in L} l_e(\lambda_{I^{(e)}_1}+\lambda_{I^{(e)}_2}),
$$
where $l_e$ is the length of the $e$th edge and $I^{(e)}_1$ and $I^{(e)}_2$ are the indices of two
(possibly coinciding) cycles incident to the $e$th edge, we obtain that
\beq
\sum_{\{P_I\}\in {\mathbb Z}_+^{s}}N_{g,s}(P_1,\dots,P_s)e^{-\sum_{I=1}^s P_I\lambda_I}
=\sum_{\Gamma_{g,s}}\frac{1}{|\hbox{Aut\,}\Gamma_{g,s}|}\prod_{e=1}^{|L|}
\frac{1}{e^{\lambda_{I^{(e)}_1}+\lambda_{I^{(e)}_2}}-1}.
\label{gen-fun-Norbury}
\eeq
We recognize in (\ref{gen-fun-Norbury}) the genus expansion of the KPMM (\ref{KPMM-1}).  We thus immediately come to the following statement:
\begin{lemma}\label{lm:Norbury1}
The generating function for the Laplace transformed \dvol\  $N_{g,s}(P_1,\dots,P_s)$ is the
KPMM (\ref{KPMM-1}). The correspondence (\ref{gen-fun-Norbury})  is given by the formula
\beq
e^{\sum'_{g,s,P_j\in {\mathbb Z}_+}N^{2-2g}\alpha^{2-2g-s} N_{g,s}(P_1,\dots,P_s)e^{-\sum_{I=1}^s P_I\lambda_I}}
=\frac{\int DX e^{-\alpha N\tr\bigl[\frac12 \Lambda X\Lambda X+\log(1-X)+X \bigr]}}
{\int DX e^{-\alpha N\tr\bigl[\frac12 \Lambda X\Lambda X-\frac12 X^2\bigr]}},
\label{KPMM-Catalan}
\eeq
where the sum ranges over stable curves with $2g-2+s>0$ and only positive perimeters $P_l$.
\end{lemma}

\begin{remark}\label{rm:Mulase}
The formula (\ref{KPMM-Catalan}) is valid at all values of $N$ and $\lambda_l$. Specializing it to the case $N=1$ (when
we have just an ordinary integral instead of the matrix one) and setting $\lambda_l=\lambda$, $\alpha=1/\hbar$, and $x=e^{\lambda}+e^{-\lambda}$, we obtain
$$
e^{\sum'_{g,s,P_j\in {\mathbb Z}_+}\hbar^{2g+s-2} N_{g,s}(P_1^2,\dots,P_s^2)e^{-\lambda\sum_{I=1}^s P_I}}
=\sqrt{\frac{1-e^{-2\lambda}}{\pi\hbar}}e^{ -({2\hbar})^{-1}e^{2\lambda}+\hbar^{-1}\lambda}F(\hbar,x),
$$
where the function
$$
F(\hbar,x):= \int_{-\infty}^\infty dt\, e^{-\frac{1}{\hbar}(t^2/2+xt+\log t)}
$$
satisfies the second-order differential equation
$$
\Bigl[\hbar^2 \frac{\partial^2 }{\partial x^2}+x \hbar\frac{\partial }{\partial x}+(1-\hbar)\Bigr] F(\hbar,x)=0.
$$
We thus reproduce the equation of the quantum curve from \cite{DM14}.
\end{remark}

\begin{remark}\label{rm:Euler}
A formal limit $P_I\to 0$ in the Laplace transformed expressions correspond to $\lambda_I\to -\infty$ in (\ref{KPMM-Catalan});
in this limit, we obtain that $\Lambda:=\hbox{diag\,}(\{e^{\lambda_I}\})\to 0$, and $N_{g,s}(0,\dots,0)$ becomes the $(g,s)$-expansion term of the Penner matrix model 
\cite{Penner} 
$$
\frac{\int DX \, e^{-\alpha N\tr \bigl[\log(1-X)+X\bigr]}}{\int DX \, e^{-\alpha N\tr X^2/2}}
$$
counting virtual Euler characteristics of moduli spaces (at negative $\alpha$),
i.e., $N_{g,s}(0,\dots,0)=(-1)^s \chi({\mathcal M}_{g,s})$.
\end{remark}

Note that the \dvol\  are quasi-polynomials: their coefficients depend on the mutual parities of the
$P_I$'s and we present one more proof of this fact below (see Corollary~\ref{cor:polynom}).
Because the corresponding generating function (\ref{KPMM-1})
is in turn (see formula (\ref{loopmeans-KPMM}))
related to the standard $s$-loop Gaussian means, or correlation functions $W_s^{(g)}$, we have the following lemma.

\begin{lemma}\label{lm:Norbury}
The correlation functions $W_s^{(g)}(x_1,\dots,x_s)$ of the Gaussian matrix model subject to the
standard topological recursion based on
the spectral curve $x=e^{\lambda}+e^{-\lambda}$, $y=\frac{1}{2}(e^{\lambda}-e^{-\lambda})$ are related to
the \dvol\  by the following explicit relation:
\beq
W_s^{(g)}(e^{\lambda_1}+e^{-\lambda_1},\dots,e^{\lambda_s}+e^{-\lambda_s})
=\prod_{I=1}^s \left[\frac{1}{e^{\lambda_I}-e^{-\lambda_I}}
\sum_{P_I=1}^\infty P_Ie^{-P_I\lambda_I}\right] N_{g,s}(P_1,\dots,P_s).
\label{Norbury-rel}
\eeq
\end{lemma}

\begin{example}\label{ex:Ms1}
We begin by considering $N_{g,1}(P)$, which are polynomials of degree $3g-2$ in $P^2$ (half the real dimension of the highest
dimensional cells in the moduli space ${\mathcal M}_{g,1}$), are nonzero only for even $P$, and must vanish
for all $P=2,\dots,4g-2$ (because the minimum number of edges of the genus $g$ shape with one face is $2g$, and
the minimum nonzero $P$ is therefore $4g$).
We thus have that, for even $P$,
$$
N_{g,1}(P)=\left(\sum_{i=0}^{g-1}q_{i}^{(g)}P^{2i}\right)\prod_{k=1}^{2g-1}\bigr(P^2-(2k)^2\bigl),
$$
and it vanishes for odd $P$, so,
after a lower-triangular change of variables $\{q_{i}^{(g)}\}\to \{b_i^{(g)}\}$ and introduction of a normalization
factor,
its Laplace transform in formula (\ref{Norbury-rel}) takes the form
\beq
W_1^{(g)}(e^{\lambda}+e^{-\lambda})=\frac{-1}{e^{\lambda}-e^{-\lambda}}
\sum_{i=0}^{g-1}\frac{b_{i}^{(g)}}{2^{4g+2i-1}(4g+2i-1)!}
\prod_{k=1}^{2g+i-1}\left(\frac{\pa^{2}}{\pa \lambda^{2}}-(2k)^2\right)\frac{\pa}{\pa \lambda}\frac{1}{e^{2\lambda}-1}.
\label{W1-1}
\eeq
We then use that
$$
-\frac{\pa}{\pa \lambda}\frac{1}{e^{2\lambda}-1}
=\frac{2}{(e^{\lambda}-e^{-\lambda})^2}
$$
and that
$$
\left(\frac{\pa^{2}}{\pa \lambda^{2}}-(2k)^2\right)\frac{1}{(e^{\lambda}-e^{-\lambda})^{2k}}
=\frac{4(2k)(2k+1)}{(e^{\lambda}-e^{-\lambda})^{2k+2}},
$$
or, in the general form,
\beq
\left(\frac{\pa^{2}}{\pa \lambda^{2}}-(m)^2\right)\frac{1}{(e^{\lambda}-e^{-\lambda})^{m}}
=\frac{4(m)(m+1)}{(e^{\lambda}-e^{-\lambda})^{m+2}},\quad m\ge1.
\label{recursion}
\eeq

Now consecutively acting by quadratic differentials in the product, we come to the general representation for
the one-loop mean,
$$
W_1^{(g)}(e^{\lambda}+e^{-\lambda})=\frac{1}{e^{\lambda}-e^{-\lambda}}
\sum_{i=0}^{g-1}b_{i}^{(g)}
\frac{1}{(e^{\lambda}-e^{-\lambda})^{4g+2i}}=\frac{1}{(e^{\lambda}-e^{-\lambda})^{4g+1}}
\sum_{i=0}^{g-1}\frac{b_{i}^{(g)}}{(e^{\lambda}-e^{-\lambda})^{2i}}.
$$
\end{example}

In Sec.~\ref{s:CohFT}, we identify the coefficients $b_{i}^{(g)}$ with ancestor invariants of a cohomological field theory.


\begin{remark}\label{rm:Pg1}
The quantities $C_{g,s}(z)$ \cite{ACRPS} are defined to be the means
from (\ref{det-rep}) weighted by $z^{\sum k_i/2}$,
$$
C_{g,s}(z):=\sum_{\{k_i\}\in {\mathbb Z}_+^s} z^{\sum k_i/2}\left\langle\prod_{i=1}^s(\tr H^{k_i})\right\rangle_g^{\mathrm{conn}}.
$$
From this form and the resolvent representation (\ref{resolvent}) it follows immediately that $C_{g,1}(z)$ is
related to the one-loop mean $W_1^{(g)}(e^\lambda+e^{-\lambda})$ by the simple
formula 
\be
W_1^{(g)}(e^\lambda+e^{-\lambda})=(e^\lambda+e^{-\lambda})^{-1} C_{g,1}\bigl((e^\lambda+e^{-\lambda})^{-2}\bigr).
\label{WC}
\ee
The statement of Corollary 1.5 of \cite{ACRPS} is that  
$$
C_{g,1}(z)=P_g(z)(1-4z)^{-3g+1/2}\ \hbox{where} \  P_g(z)=z^{2g}\sum_{k=0}^{g-1}
P^{(g)}_k z^k,
$$ 
and $P^{(g)}_k$ are positive integers. The explicit relation between $P^{(g)}_k$ and $b_{k}^{(g)}$ is as follows. If we 
substitute $z=(e^\lambda+e^{-\lambda})^{-2}$ into the above relation and use (\ref{WC}), we obtain that
\be
W_1^{(g)}(e^\lambda+e^{-\lambda})=\sum_{k=0}^{g-1}P^{(g)}_k \frac{(e^\lambda+e^{-\lambda})^{2k}}{(e^\lambda-e^{-\lambda})^{6g-1}}.
\label{WC2}
\ee

A more general relation, which follows from Lemma~\ref{lm:multiloop} upon equating all variables $\lambda_I$, reads
\be
W_s^{(g)}(e^\lambda+e^{-\lambda},\dots,e^\lambda+e^{-\lambda})
=\sum_{k=0}^{g+s-2}P^{(g,s)}_k \frac{(e^\lambda+e^{-\lambda})^{2k}}{(e^\lambda-e^{-\lambda})^{6g+5s-6}},
\ee
and generating functions $C_{g,s}(z)$ for $s$-backbone $g$-genus chord diagrams are thus determined by $g+s-1$ coefficients 
$P^{(g,s)}_k $. 

Turning back to the one-backbone case, note that, because $(e^\lambda+e^{-\lambda})^2=(e^\lambda-e^{-\lambda})^2+4$, we can express $P^{(g)}_k$ through $b_{k}^{(g)}$ and vice versa:
\bea
&{}&b^{(g)}_{g-1-k}=\sum_{l=k}^{g-1}P^{(g)}_l 4^{l-k}\left({l\atop k}\right),\nonumber\\
&{}&P^{(g)}_{k}=\sum_{l=k}^{g-1}b^{(g)}_{g-1-l} (-4)^{l-k}\left({l\atop k}\right),\nonumber
\eea
and, in particular, the integrality of the $P^{(g)}_{k}$ proved in \cite{ACRPS} implies that of $b^{(g)}_{k}$ and vice versa. We prove the
positivity and integrality of $b_{k}^{(g)}$ in Lemma~\ref{lm2} below; this does not however imply the positivity of $P^{(g)}_{k}$ observed
and conjectured in \cite{ACRPS}. 
\end{remark}

\subsection{The Kontsevich--Penner matrix model}
\label{ss:discretemoduli}

The matrix model (\ref{KPMM-1}) manifests many remarkable
properties:
\begin{enumerate}
\item It is equivalent~\cite{ChMak2,KMMMZ} to the Hermitian matrix model with the potential determined by the Miwa change of the variables $t_k=\frac{1}{k}\tr (e^\Lambda+e^{-\Lambda})^{-k}+\frac12 \delta_{k,2}$;
\item In the special times $T^{\pm}_{2r}$, $r=0,1,\dots$, (\ref{times-KP}),
which is related to the {\em discretization of moduli spaces}, is equal to the product of two Kontsevich matrix
models \cite{Ch95}, intertwined by a canonical transformation of the variables
(see Lemma~\ref{lm:canonical});
\item It is the generating function for the \dvol\  $N_{g,s}(P_1,\dots,P_s)$;
\item It is related to the multiloop correlation functions of the Gaussian model;
\item It is the generating function for the number of clean Belyi functions, or for the
corresponding Grothendieck {\em dessins d'enfant} \cite{AmCh14} (see also \cite{AMMN}).
\end{enumerate}

As concerning the second of the above properties, we have the following exact relation.

\begin{lemma}\label{lm:canonical}(\cite{Ch95})
The partition function ${\mathcal Z}[\Lambda]$ (\ref{KPMM-1})
expressed in the times $T^\pm_{k}(\lambda)$ (\ref{times-KP}) depends only on the even times $T^\pm_{2k}(\lambda)$
and satisfies the following exact relation:
\beq
{\mathcal Z}[\Lambda]=e^{{\mathcal F}_{\text{KP}}[\{T^\pm_{2n}\}]}
=e^{C(\alpha N)}e^{-N^{-2}{\mathcal A}}e^{{\mathcal F}_{\text{K}}[\{T^+_{2n}\}]+{\mathcal F}_{\text{K}}[\{T^-_{2n}\}]},
\label{canonical}
\eeq
where ${\mathcal F}_{\text{K}}[\{T^\pm_{2n}\}]$ is a free energy of the
Kontsevich matrix model (\ref{Kont}) with $T^\pm_{2n}$ being the times of the KdV hierarchies and ${\mathcal A}$ is the operator of a canonical transformation,
\bea
{\mathcal A}&=&\sum_{m,n=0}^\infty \frac{B_{2(n+m+1)}}{4(n+m+1)}\frac{1}{(2n+1)!(2m+1)!}
\Bigl\{\frac{\pa^2}{\pa T^+_{2n}\pa T^+_{2m}}+\frac{\pa^2}{\pa T^-_{2n}\pa T^-_{2m}}+
2(2^{2(n+m+1)}-1)\frac{\pa^2}{\pa T^+_{2n}\pa T^-_{2m}}\Bigr\}\nonumber\\
&{}&+\sum_{n=2}^\infty \alpha N^2\frac{2^{2n-1}}{(2n+1)!}\Bigl(\frac{\pa}{\pa T^-_{2n}}+\frac{\pa}{\pa T^+_{2n}} \Bigr).
\label{A}
\eea
Here $C(\alpha N)$ is a function depending only on $\alpha N$ that ensures that
${\mathcal F}_{\text{KP}}[\{T^\pm_{2n}\}]=0$ for $T^\pm_{2n}\equiv 0$ and $B_{2k}$ are the Bernoulli numbers
generated by $t/(e^t-1)=\sum_{m=0}^\infty B_mt^m/(m!)$ and $T^\pm_{2k}$ are given by (\ref{times-KP}).
\end{lemma}

From this canonical transformation we immediately obtain the (ordinary) graph representation for the
term ${\mathcal F}_{g,s}[\{T^\pm_{2n}\}]$ of the expansion of
$$
{\mathcal F}_{\text{KP}}[\{T^\pm_{2n}\}]=\sum_{g,s} N^{2-2g}\alpha^{2-2g-s} {\mathcal F}_{g,s}[\{T^\pm_{2n}\}].
$$

\begin{lemma}\label{lm:graph}
The term ${\mathcal F}_{g,s}[\{T^\pm_{2n}\}]$ of the genus expansion of the KPMM (\ref{KPMM-1}) is given by 
a sum of the finite set of graphs $G_{g,s}$ described below, with each graph contributing the factor also described below decided by the order of the automorphism group of the graph. 
\begin{itemize}
\item each vertex $v_i$, $i=1,\dots,q$, of
a graph $G_{g,s}$ carries the marking "$+$" or "$-$", genus $g_i\ge 0$, and has $s_i$
endpoints of edges
incident to it ($2g_i-2+s_i>0$, i.e., all vertices are stable);
each endpoint of an edge carries a nonnegative integer $k^{\pm}_{r,i}$, $r=1,\dots,s_i$,
where the superscript $+$ or $-$ is determined by the marking of the vertex;
\item edges can be external legs (ordinary leaves) with $k^{\pm}_{r,i}\ge 0$ (we let $a_i\ge 0$ denote the number
of such legs incident to the $i$th vertex),
half-edges (dilaton leaves) with $k^{\pm}_{r,i}\ge 2$ (we let $b_i\ge 0$ denote the number
of such legs incident to the $i$th vertex), or internal edges incident either to two different vertices or to the same vertex (their two endpoints carry in general different numbers $k^{\pm}_{r_1,i_1}$ and $k^{\pm}_{r_2,i_2}$) (we let $l_i$ denote the number of internal edge endpoints
incident to the $i$th vertex);
\item each vertex contributes the Kontsevich intersection index
$$
\Bigl\langle \tau_{k^{\pm}_{1,i}}\cdots \tau_{k^{\pm}_{s_i,i}}\Bigr\rangle_{g_i}
$$
with $\sum_{r=1}^{s_i}k^{\pm}_{r,i}=3g_i-3+s_i$;
\item every internal edge with endpoint markings $(k_1^{+},k_2^{+})$ or $(k_1^{-},k_2^{-})$
(two endpoints of such an edge can be incident to the same vertex)
contribute the factor
$$
-\frac{B_{2(k^{\pm}_1+k^{\pm}_2+1)}}{2(k^{\pm}_1+k^{\pm}_2+1)}
\frac{1}{(2k^{\pm}_1+1)!(2k^{\pm}_2+1)!}
$$
and every internal edge with endpoint markings $(k_1^{+},k_2^{-})$
(two endpoints of such an edge can be incident only to distinct vertices with different markings $+$ and $-$)
contributes the factor
$$
-\frac{B_{2(k^+_1+k^-_2+1)}}{2(k^+_1+k^-_2+1)}
\frac{2^{2(k^+_1+k^-_2+1)}-1}{(2k^+_1+1)!(2k^-_2+1)!};
$$
\item every half-edge with the marking $r^{\pm}\ge 2$ carries the factor $-\frac{2^{2r^{\pm}-1}}{(2r^{\pm}+1)!}$;
\item every external leg with the marking $k^{\pm}_{r,i}$ contribute the corresponding time
$T^{\pm}_{2k^{\pm}_{r,i}}$;
\item $\sum_{i=1}^q (g_i+l_i/2-1)+1=g$ (the total genus $g$ is equal to the sum of internal genera plus the
number of loops in the graph);
\item $\sum_{i=1}^q a_i=s$ (the total number of external legs is fixed and equal to $s$);
\end{itemize}
From the above formulas, we have that
\beq
\sum_{j=1}^s k_j^{\text{Ext}}=3g-3+s-\sum_{j=1}^{|L|}(1+k^{\text{Int}}_{j,1}++k^{\text{Int}}_{j,2})
-\sum_{j=1}^{|B|}(k^{\text{Half}}_j-1),
\label{reducing}
\eeq
where, disregarding the vertex labels,
$k_j^{\text{Ext}}\ge 0$ are indices of the external edges, $k^{\text{Int}}_{j,1}\ge 0$ and $k^{\text{Int}}_{j,2}\ge 0$
are indices of endpoints of the internal edges, $k^{\text{Half}}_j\ge 2$ are indices of half-edges, and $|L|$ and
$|B|$ are the cardinalities of the respective sets of internal edges and half-edges of the graph.
\end{lemma}

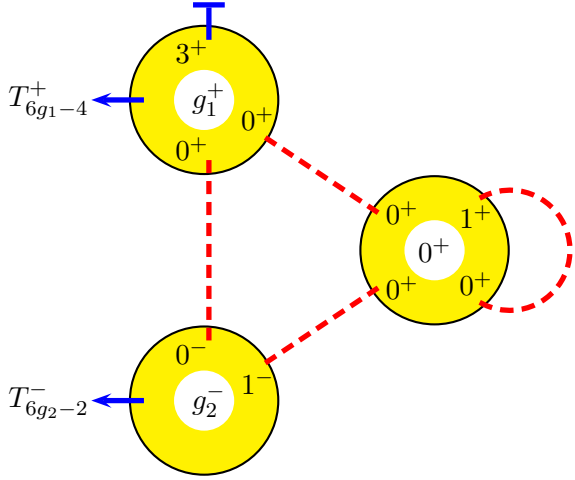
\begin{figure}[h]
{\psset{unit=1}
\begin{pspicture}(-7,-3.6)(5,3.6)
\newcommand{\PATTERN}{%
{\psset{unit=1}
\pscircle[fillstyle=solid,fillcolor=yellow](0,0){1}
\pscircle[fillstyle=solid,fillcolor=white,linecolor=white](0,0){0.4}
\psline[linewidth=2pt,linecolor=blue]{->}(-0.8,0)(-1.5,0)
}
}
\rput(-2,2){\PATTERN}
\rput(-3.6,2){\makebox(0,0)[rc]{$T^+_{6g_1-4}$}}
\rput(-2,-2){\PATTERN}
\rput(-3.6,-2){\makebox(0,0)[rc]{$T^-_{6g_2-2}$}}
\psline[linewidth=2pt,linecolor=blue]{-|}(-2,2.8)(-2,3.3)
\rput(-2,2.8){\makebox(0,0)[rt]{$3^+$}}
\pscircle[fillstyle=solid,fillcolor=yellow](1,0){1}
\pscircle[fillstyle=solid,fillcolor=white,linecolor=white](1,0){0.4}
\rput(1.55,0.6){\makebox(0,0)[ct]{$1^+$}}
\rput(1.55,-0.6){\makebox(0,0)[cb]{$0^+$}}
\rput(-2,2){\makebox(0,0)[cc]{$g_1^+$}}
\rput(-2,-2){\makebox(0,0)[cc]{$g_2^-$}}
\rput(1,0){\makebox(0,0)[cc]{$0^+$}}
\psarc[linecolor=red,linestyle=dashed,linewidth=2pt](2,0){0.8}{-120}{120}
\psline[linecolor=red,linestyle=dashed,linewidth=2pt](-2,1.2)(-2,-1.2)
\rput(-2,1.2){\makebox(0,0)[rb]{$0^+$}}
\rput(-2,-1.2){\makebox(0,0)[rt]{$0^-$}}
\psline[linecolor=red,linestyle=dashed,linewidth=2pt](-1.25,1.5)(0.25,0.5)
\rput(-1.35,1.6){\makebox(0,0)[cb]{$0^+$}}
\rput(0.35,0.5){\makebox(0,0)[lc]{$0^+$}}
\psline[linecolor=red,linestyle=dashed,linewidth=2pt](-1.25,-1.5)(0.25,-0.5)
\rput(-1.35,-1.6){\makebox(0,0)[ct]{$1^-$}}
\rput(0.35,-0.5){\makebox(0,0)[lc]{$0^+$}}
\end{pspicture} }
\caption{\small The typical diagram from the graph expansion $G_{g,s}$: it contains three vertices with labels $(g_1,+)$, $(g_2,-)$, 
$(0,+)$ and two loops, so the total  genus is $g=g_1+g_2+0+2$. The number of external legs is two, so the highest possible total degree of $T^{\pm}$ is $6g-6+4=6g_1+6g_2+10$. The actual total degree is $6g_1+6g_2-6$ and it corresponds to a reduction of level eight. Recall that every dilaton leaf with marking $k$ reduces the total degree in times by $2[k-1]$, every insertion of an edge with markings
$(k_1^\pm,k_2^\pm)$ reduces the total degree by $2[k_1+k_2+1]$. We have four edges and one dilaton leaf, the total reduction of degree
is $2[3-1]+2[0+0+1]+2[0+0+1]+2[0+1+1]+2[0+1+1]=16$, as expected.} 
\label{fi:primer}
\end{figure}

The \emph{proof} is just another application of Wick's theorem, now in the form of exponential of a 
linear-quadratic differential operator: when acting on the exponential of any combination of variables,
the quadratic part of (any)
linear-quadratic differential operator
defines the pairing; the sum over all possible pairings of an exponentiated quantity (which is in our case the sum of free energies of two Kontsevich models) is in turn the exponential of the sum over all \emph{connected} pairings of these quantities. The linear part of the differential operator just produces constant shifts of the higher times, which can also be described by insertions of half-edges. We present a typical term in the resulting free-energy expansion in Fig.~\ref{fi:primer}.

This lemma immediately implies the corollary
\begin{corollary}\label{cor:polynom}
The quantities ${\mathcal F}_{g,s}[\{T^\pm_{2n}\}]$ are polynomials in the times such that, for every monomial
$T^+_{2n_1}\cdots T^-_{2n_s}$ we have that $\sum_{i=1}^s n_i\le 3g-3+s$, and the highest term
with $\sum_{i=1}^s n_i= 3g-3+s$ is
$$
\bigl\<\tau_{n_1}\cdots \tau_{n_s}\bigr\>_g\Bigl(\prod_{i=1}^s T^+_{2n_i}+\prod_{i=1}^s T^-_{2n_i}\Bigr).
$$
This also implies that all \dvol\  $N_{g,s}(P_1,\dots,P_s)$ are ${\mathbb Z}_2$-quasi-polynomials in $P^2_I$.
\end{corollary}

\noindent
{\bf Proof.}  The \dvol\  $N_{g,s}(P_1,\dots,P_s)$
depend only on even powers of $P_I$ because ${\mathcal F}_{g,s}$ depend only on even times
$T^\pm_{2n}$; the quasi-polynomiality follows immediately from the fact that ${\mathcal F}_{g,s}$ are
polynomials in $T^\pm_{2n}$.

\begin{remark}\label{rm:alternative}
Note that the quadratic part of the differential operator (\ref{A}) matches the alternating structure of (\ref{conj}). Indeed,
the Bernoulli numbers $B_{2n}$ manifest an alternating-sign structure, $B_{2n}$ are positive for odd $n$ and negative for
even $n$,
$$
B_{2n}=(-1)^{n+1} \frac{2(2n)!}{(2\pi)^{2n}}\Bigl[1+\frac{1}{2^{2n}}+\frac{1}{3^{2n}}+\cdots\Bigr].
$$
Every insertion of an edge with a contributing $B_{2n}$ decreases the total power $k_1+\cdots+k_s$ of the corresponding monomial
$T^{\pm}_{2k_1}\cdots T^{\pm}_{2k_s}$ exactly by $n$. With the overall minus sign in front of
${\mathcal A}$ in (\ref{A}) taken into account, this gives the desired alternating-sign structure of (\ref{conj}) up to constant shifts of the higher times
$T^\pm_{2k}$ with $k\ge 2$ ensured by the dilaton leaves.
\end{remark}

\subsection{The times for the multi-resolvents}\label{ss:multi-times}

We now consider the general $s$-resolvent case.
We reformulate Lemma~\ref{lm:graph} as the following statement

\begin{lemma}\label{lm-polynomiality}
From (\ref{loopmeans-KPMM}) we have that
the (stable) loop means (with $2g+s-2\ge1$) are polynomials in the times for any genus $g$ and number of loops $s$:
\bea
&{}&W_s^{(g)}(e^{\lambda_1}+e^{-\lambda_1},\dots,e^{\lambda_s}+e^{-\lambda_s})
=F_{g,s}\bigl(\{t^{\pm}_{2n_j+1}(\lambda_{j})\}\bigr),
\label{main}
\eea
where we have to resubstitute the times $T^{\pm}_{2d}$ for the {\em new times}
\beq
\label{times}
T^{\pm}_{2d}\to t^{\pm}_{2d+1}(\lambda_{j}):=
\frac{1}{e^{\lambda_j}-e^{-\lambda_j}}
\left(\frac{\partial}{\partial \lambda_j}\right)^{2d+1}\frac{1}{e^{\lambda_i}\pm 1},
\eeq
which are derivatives of the times $T^{\pm}_{2d}$.
Note that all the times $t^{\pm}_{2d+1}(\lambda)$ are now strictly
skew-symmetric with respect to the change of variables $\lambda\to -\lambda$.

The two nonstable loop means are
\bea
&{}&W_1^{(0)}(e^{\lambda}+e^{-\lambda})=e^{-\lambda},
\label{W10}\\
&{}&W_2^{(0)}(e^{\lambda_1}+e^{-\lambda_1},e^{\lambda_2}+e^{-\lambda_2})=\prod_{i=1,2}\prod_{j=1,2}
\frac{1}{e^{\lambda_i}-e^{-\lambda_j}}
\label{W20}
\eea
\end{lemma}

The \emph{proof} is the observation that the stable loop means are related to the terms of the free-energy expansion of the
KPMM by (\ref{loopmeans-KPMM}) and the action of the differential operators in this formula makes the time change (\ref{times}). 
The first nonstable loop mean $W_1^{(0)}$ comes just from the Catalan number counting and the second one, $W_2^{(0)}$, comes from
the fact that $F_{0,2}(\lambda_1,\lambda_2)=-\log\bigl(1-e^{-\lambda_1-\lambda_2} \bigr)$ is the term from the normalisation factor in
the KPMM and acting by 
$\frac{1}{e^{\lambda_1}-e^{-\lambda_1}}\frac{\pa}{\pa \lambda_1}\frac{1}{e^{\lambda_2}-e^{-\lambda_2}}\frac{\pa}{\pa \lambda_2}$ on it we obtain (\ref{W20}).

We obtain another useful set of times exploiting formula (\ref{W1-1}) from Example~\ref{ex:Ms1}. First, we obviusly have that
\beq
\label{time-w}
t^{-}_{2d+1}(\lambda)+t^{+}_{2d+1}(\lambda)=
\frac{1}{e^{\lambda}-e^{-\lambda}}
\left(\frac{\partial}{\partial \lambda}\right)^{2d+1}\frac{2}{e^{2\lambda}-1}
=\sum_{j=1}^{d+1}q_{j,d}\frac{1}{(e^\lambda-e^{-\lambda})^{2j+1}}
\eeq
and
\beq
\label{time-as}
t^{-}_{2d+1}(\lambda)-t^{+}_{2d+1}(\lambda)=
\frac{1}{e^{\lambda}-e^{-\lambda}}
\left(\frac{\partial}{\partial \lambda}\right)^{2d+1}\frac{2}{e^{\lambda}-e^{-\lambda}}
=\sum_{j=1}^{d+1}\tilde q_{j,d}\frac{e^\lambda+e^{-\lambda}}{(e^\lambda-e^{-\lambda})^{2j+1}}
\eeq
with some integer coefficients $q_{j,d}$ and $\tilde q_{j,d}$.
The relation (\ref{time-as}) follows from
$$
\frac{1}{e^\lambda-1}+\frac{1}{e^\lambda+1}=\frac{2}{e^\lambda-e^{-\lambda}}
$$
and from another useful representation:
\bea§
\frac{1}{e^{\lambda}-e^{-\lambda}}\frac{\partial}{\partial\lambda}
\prod_{k=1}^{d}\left(\frac{\pa^{2}}{\pa \lambda^{2}}-(2k-1)^2\right)\frac{2}{e^\lambda-e^{-\lambda}}
&=&\frac{1}{e^{\lambda}-e^{-\lambda}}\frac{\partial}{\partial\lambda}
\frac{2^{2d+1} (2d)!}{(e^\lambda-e^{-\lambda})^{2d+1}}\nonumber\\
&=&-2^{2d+1}(2d+1)!
\frac{e^\lambda+e^{-\lambda}}{(e^\lambda-e^{-\lambda})^{2d+3}}
\label{W1-as}
\eea
We can therefore equivalently expand $F_{g,s}\bigl(\{t^{\pm}_{2n_j+1}(\lambda_{j})\}\bigr)$
in  the variables
\beq
\label{sj}
s_{k,\beta}(\lambda):=\frac{(e^{\lambda}+e^{-\lambda})^\beta_{}}{{(e^{\lambda}-e^{-\lambda})^{2k+3}}},
\quad k=0,\dots,3g+s-3, \ \beta=0,1.
\eeq
In the next section we demonstrate that the coefficients of these expansions
are related to the ancestor invariants of a CohFT.

We now present the general structure of the stable multiloop mean.
\begin{lemma}\label{lm:multiloop}
The general expression for a stable ($2g+s-3\ge 0$) loop mean
$W_s^{(g)}(e^{\lambda_1}+e^{-\lambda_1},\dots,e^{\lambda_s}+e^{-\lambda_s})$ in terms of
the variables $s_{k,\beta}(\lambda)$ given by (\ref{sj}) reads:
\beq
W_s^{(g)}(e^{\lambda_1}+e^{-\lambda_1},\dots,e^{\lambda_s}+e^{-\lambda_s})
=\sum_{\vec{k},\vec{\beta}}\widehat{b}^{(g)}_{\vec{k},\vec{\beta}}\prod_{j=1}^s s_{k_j,\beta_j}(\lambda_j),
\label{W->sk}
\eeq
where $k_j$ and $\beta_j$ are subject to the restrictions:
\beq
2g-1+\frac12\sum_{j=1}^s\beta_j\le \sum_{j=1}^s k_j\le 3g+s-3,
\qquad \sum_{j=1}^s\beta_j=0\ \hbox{mod}\ 2.
\label{restr-k}
\eeq
\end{lemma}

The {\em proof} for the restrictions follows from two considerations: first, as $\lambda_j\to \infty$ uniformly for
all $j$, that is, we scale $\lambda_j\to \lambda_j+R$, every edge contributes a factor $e^{-2R}$ plus $s$ factors
$e^{-R}$ due to the derivatives. The minimum number of edges (for a shape with one vertex) is $2g+s-1$, so the minimum
factor appearing is $e^{(-4g-3s+2)R}$ whereas $s_{k,\beta}(\lambda)$ scale as $e^{(-3-2k+\beta)R}$, which results
in the lower estimate. The upper estimate follows from the analysis of the pole behaviour as $\lambda_j = 0$. On the one hand, $s_{k,\beta}(\lambda)\sim \lambda^{-2k-3}$ as $\lambda\to 0$ irrespectively of $\beta$; on the other hand,
from the relation to the Kontsevich model we can conclude that the pole structure of the derivatives of the Kontsevich KdV times is $t_{d_j}(\lambda_j)\sim \lambda_j^{-2d_j-3}$ with $\sum_j d_j\le 3g+s-3$ and therefore
$\sum_j d_j=\sum_j k_j$, which leads to the upper estimate. That the sum of the $\beta_j$'s is even follows from
the fact that the total expression is symmetric with respect to the total change of the times $T^{\pm}\to T^{\mp}$; under this change,
the variables $s_{k,\beta}(\lambda)$ behave as $s_{k,\beta}(\lambda)\to (-1)^\beta s_{k,\beta}(\lambda)$,
so the sum of the beta factors must be even.

\section{Cohomological field theory from \dvol}\label{s:CohFT}

In this section we describe a cohomological field theory (CohFT) associated to the \dvol.  A dimension $d$ Frobenius manifold structure is equivalent to a CohFT  for a dimension $d$ vector space $H$ with basis $\{e_{\alpha}\}$ and metric $\eta$.  The main result is Theorem~\ref{th:cohft} which shows that the quasi-polynomial \dvol\  are equivalent to the correlators in the CohFT associated to the Hurwitz Frobenius manifold $H_{0,(1,1)}$ described in the introduction.  We give two proofs of the genus 0 case of this, because one of the proofs is constructive and the other generalises to all genus.  The constructive proof gives an explicit realisation of the genus 0 CohFT, and it proves that the CohFT satisfies good properties---in particular it is a homogenous CohFT.  The homogeneity condition goes some way towards explaining the appearance of $\chi(\modm_{g,n})$ as the primary correlators in the CohFT.

\subsection{Cohomological field theories}\label{ss:Frob}

Given a complex vector space $H$ equipped with a complex metric $\eta$, a CohFT is a sequence of $S_s$-equivariant linear maps
\[ I_{g,s}:H^{\otimes s}\to H^*(\overline{\modm}_{g,s})\]
which satisfy the following compatibility conditions with respect to inclusion of strata.
Any partition into two disjoint subsets $I\sqcup J=\{1,\dots,s\}$ defines a map
$$\phi_I:\overline{\modm}_{g_1,|I|+1}\times\overline{\modm}_{g_2,|J|+1}\to\overline{\modm}_{g,s}$$
and
$$\phi_I^*I_{g,s}(v_1\otimes\cdots\otimes v_s)=I_{g_1,|I|+1}\otimes I_{g_2,|J|+1}\left(\bigotimes_{i\in I}v_i\otimes\Delta\otimes\bigotimes_{j\in J}v_j\right)$$
where $\Delta=\sum_{\alpha,\beta}\eta^{\alpha\beta}e_{\alpha}\otimes e_{\beta}$ with respect to a basis $\{e_{\alpha}\}$ of $H$.  The map
$$\psi:\overline{\modm}_{g-1,s+2}\to\overline{\modm}_{g,s}$$
induces
$$\psi^*I_{g,s}(v_1\otimes\cdots\otimes v_s)=I_{g-1,s+2}(v_1\otimes\cdots\otimes v_s\otimes\Delta).$$
The three-point function $I_{0,3}$ together with the metric $\eta$ induces a product $\bullet$ on $H$:
$$u\bullet v=\sum_{\alpha,\beta}I_{0,3}(u\otimes v\otimes e_{\alpha})\eta^{\alpha\beta}e_{\beta}$$
where $I_{0,3}$ takes its values in $\bc$.   A vector $e_0$ satisfying
$$I_{0,3}(v_1\otimes v_2\otimes e_0)=\eta(v_1\otimes v_2),\quad \forall v_1,v_2\in H
$$
is the identity element for the product on $H$.

One further condition, which is not an axiom for a general CohFT, is satisfied by the CohFT we consider here (and more generally, for example it is satisfied by Gromov--Witten invariants).   For $s\geq 3$, the forgetful map
$$\pi:\overline{\modm}_{g,s+1}\to\overline{\modm}_{g,s}$$
induces
$$
I_{g,s+1}(v_1\otimes\cdots\otimes v_s\otimes e_0)=\pi^*I_{g,s}(v_1\otimes\cdots\otimes v_s).
$$

\subsection{Quasipolynomials and ancestor invariants}\label{ss:quasi}
The \dvol\  $N_{g,s}(P_1,\dots,P_s)$ are mod 2 even quasi-polynomials, i.e. it is an even polynomial on each coset of $2\bz^s\subset\bz^s$.   Define a basis of mod 2 even quasi-polynomials induced (via tensor product) from the following single-variable basis $p_{k,\alpha}(b)$ for $k=0,1,2,\dots$ and $\alpha=0,1$.
$$p_{0,0}(b)=\left\{\begin{array}{cc}1,&b \text{\ even}\\0,&b \text{\ odd}\end{array}\right.,\quad p_{0,1}(b)=\left\{\begin{array}{cc}0,&b \text{\ even}\\1,&b \text{\ odd}\end{array}\right.$$
$$p_{k+1,\alpha}(b)=\sum_{m=0}^b m p_{k,\alpha}(m),\quad k\geq 0.$$
$$p_{k,\alpha}(b)=\frac{p_{0,k+\alpha}(b)}{4^kk!}\hspace{-.5cm}\prod_{\begin{array}{c}0<m\leq k\\ m\equiv k+\alpha\ \text{(mod 2)}\end{array}}\hspace{-.8cm}(b^2-m^2)$$
where in the second subscript we mean $k+\alpha$ (mod 2).

Put $\vec{k}=(k_1,\dots,k_s)$ and  $\vec{\alpha}=(\alpha_1,\dots,\alpha_s)$.
\begin{theorem} \label{th:cohft} We have that
$$N_{g,s}(P_1,\dots,P_s)=\sum_{\vec{k},\vec{\alpha}} c^g_{\vec{k},\vec{\alpha}} \prod_{i=1}^s p_{k_i,\alpha_i}(P_i)$$
where the coefficients are ancestor invariants:
\be
c^g_{\vec{k},\vec{\alpha}}=\int_{\overline{\modm}_{g,s}}I_{g,s}(e_{\alpha'_1}\otimes\cdots\otimes e_{\alpha'_s})\prod_{i=1}^s\psi_i^{k_i}.
\label{ancestor}
\ee 
\end{theorem}
The theorem can be proven as an application of \cite{DOSSIde} which identifies theories with a special class of spectral curves with semisimple CohFTs.  We need to check that the spectral curve satisfies the necessary conditions.  The outcome of applying \cite{DOSSIde} is non-constructive so we prove the genus 0 case in a different way that enables an explicit realisation of the CohFT.

\subsection{Genus 0 reconstruction.}

The {\em primary} correlators of a CohFT are defined by:
$$Y_{g,s}:=\int_{\overline{\modm}_{g,s}}I_{g,s}:H^{\otimes s}\to\bc$$
which assemble into the generating function
$$F(t_0,...,t_{D-1})=\sum N^{2-2g}\frac{1}{s!}Y_{g,s}=\sum N^{2-2g}F_g
$$
where $(t_0,...,t_{D-1})$ in $ H^*$ is the dual basis of $\{e_0,...,e_{D-1}\}$.  The genus 0 part $F_0$ is the prepotential of the CohFT.

The genus zero primary correlators satisfy relations coming from Keel's relations in $H^*(\overline{\modm}_{0,s})$ \cite{KeeInt} which we will not write here, and instead use the fact that they are equivalent to the prepotential $F_0$ satisfying the WDVV equation.  When $\dim H<3$ (as is the case here) the WDVV equation is trivial.  Conversely, Manin defines a system of {\em abstract correlation functions} to be a collection of symmetric functions $Y_{0,s}:H^{\otimes s}\to\bc$ that form a prepotential $F_0$ satisfying the WDVV equation.

\begin{theorem}[Manin \cite{ManFro} Theorem III.4.3]   \label{th:manin}
One can uniquely reconstruct a genus 0 CohFT from abstract correlation functions.  
\end{theorem}
The Deligne--Mumford compactification $\overline{\modm}_{g,s}$ possesses a natural stratification indexed by {\em dual graphs}. The dual graph of $\Sigma\in\overline{\modm}_{g,s}$ has vertices corresponding to the irreducible components of $\Sigma$ and assigned genus, edges corresponding to the nodes of $\Sigma$, and a {\em tail}---an edge with an open end (no vertex)---corresponding to each labeled point of $\Sigma$.  If $\Gamma$ is a dual graph of type $(g,s)$, then the collection of curves $D_{\Gamma}$ whose associated dual graph is $\Gamma$ forms a stratum of $\modm_{g,s}$.  The closure $\overline{D}_{\Gamma}=\cup_{\Gamma'<\Gamma}D_{\Gamma'}$, where the partial ordering is given by edge contraction, represents an element of $H^*(\overline{\modm}_{g,s})$.   Keel \cite{KeeInt} proved that $H^*(\overline{\modm}_{0,s})$ is generated by $\overline{D}_{\Gamma}$ and gave all relations.

The proof of Theorem~\ref{th:manin} uses that
$$ \int_{\overline{D}_{\Gamma}}I_{0,s}(v_1\otimes\cdots\otimes v_s)=\bigotimes_{v\in V_{\Gamma}}Y_{0,|v|}\left(\bigotimes _{i=1}^s v_i\otimes\Delta^{\otimes |E_{\Gamma}|}\right).
$$
which defines evaluation of a cohomology class on boundary strata tautologically from the definition of a CohFT.  Since $H^*(\overline{\modm}_{0,s})$ is generated by its boundary strata, and relations in $H^*(\overline{\modm}_{0,s})$ agree with the relations satisfied by abstract correlation functions, this is enough to prove the theorem.

In particular, the primary invariants 
\bea
&{}&Y_{0,3}(e_0\otimes e_0\otimes e_1)=1=Y_{0,3}(e_1\otimes e_1\otimes e_1), \quad Y_{0,s}(e_0\otimes \hbox{anything})=0,\quad 
s>3\nonumber\\
&{}&Y_{0,s}(e_1^{\otimes s})=N_{0,s}(0,...,0)=\chi(\modm_{0,s})\quad s>3
\eea
define a genus 0 CohFT.  The claim of Theorem~\ref{th:cohft}, that the coefficients of $N_{0,s}$ are the correlators of a CohFT, is an overdetermined system, since the CohFT has been uniquely determined by the constant terms $N_{0,s}(0,...,0)$.  To prove that the coefficients of $N_{0,n}$ are indeed the correlators of this CohFT we use homogeneity of the correlation functions and the CohFT.

\subsection{A homogeneous CohFT}
A CohFT is conformal if its prepotential is quasihomogeneous with respect to a vector field $E$ known as the {\em Euler vector field}:
\begin{equation}  \label{euler}
 E\cdot F_0=(3-d)F_0+Q(t)
 \end{equation}
where $Q$ is a quadratic polynomial in $t=(t_0,...,t_{D-1})$.  Using the genus 0 reconstruction in Theorem~\ref{th:manin}, Manin proved that a conformal CohFT induces the following push-forward condition on the genus 0 CohFT.

Let $\xi$ be any vector field on $H$ which we consider to be a manifold with coordinates $t_0,...,t_{D-1}\in H^*$.   The Lie derivative with respect to $\xi$ of the correlators of a CohFT $I_{g,s}$  induces a natural action on the CohFT:
$$(\xi\cdot I)_{g,s}(v_1\otimes\cdots\otimes v_s)=\deg I_{g,s}(v_1\otimes\cdots\otimes v_s)-\sum_{j=1}^s I_{g,s}(v_1\otimes..\otimes[\xi,v_j]\otimes..\otimes v_s)+\pi_*I_{g,s+1}(v_1\otimes...\otimes v_s\otimes\xi)
$$
where $\pi:\overline{\modm}_{g,s+1}\to\overline{\modm}_{g,s}$ is the forgetful map.

Here $I$ is a ($H^*(\overline{\modm}_{0,s})$-valued) tensor on $H$ which is acted on infinitesimally by the vector field $\xi$ on $H$.

A CohFT is {\em homogeneous} of weight $d$ if
\begin{equation}  \label{homog}
(E.I)_{g,s}=((g-1)d+s)I_{g,s}
 \end{equation}
If the preprotential satisfies the homogeneity condition (\ref{euler}) then the proof of Theorem~\ref{th:manin} can be used to prove that the genus 0 CohFT is homogeneous.  The Lie derivative of the bivector $\Delta$ dual to the metric $\eta$ on $H$ can be calculated in flat coordinates
$$\cl_E\cdot\Delta=\cl_E\cdot\eta^{ij}e_i\otimes e_j=\eta^{ij}([E,e_i]\otimes e_j+e_i\otimes [E,e_j])=(d-2)\eta^{ij}e_i\otimes e_j=(d-2)\Delta
$$
where we have used a choice of flat coordinates \cite{DubGeo} with respect to which $\eta=\delta_{i,D-1-i}$ and
$$E=\sum_i(\alpha_it_i+\beta_i)\frac{\partial}{\partial t_i}
$$
where $\alpha_i+\alpha_{D-1-i}=2-d$.
 The common weight $d$ in the homogeneous conditions (\ref{euler}) and (\ref{homog}) arises from the following compatibility for pull-backs in a homogeneous CoHFT.
\begin{eqnarray*}
p^*E\cdot I_{g,s}&=&E\cdot p^*I_{g,s}=E\cdot(I_{g_1,s_1+1}\otimes I_{g_2,s_2+1}(\Delta))\\
&=&
(E\cdot I)_{g_1,s_1+1}\otimes I_{g_2,s_2+1}(\Delta)+I_{g_1,s_1+1}\otimes (E\cdot  I)_{g_2,s_2+1}(\Delta)
+I_{g_1,s_1+1}\otimes   I_{g_2,s_2+1}(\cl_E\cdot\Delta)\\
&=&
[((g_1-1)d+s_1+1)+((g_2-1)d+s_2+1)+(d-2)]I_{g_1,s_1+1}\otimes I_{g_2,s_2+1}(\Delta)\\
&=&((g-1)d+s)I_{g_1,s_1+1}\otimes I_{g_2,s_2+1}(\Delta)
\end{eqnarray*}

\subsection{Proof of Theorem~\ref{th:cohft} in genus 0.}

We are now in a position to prove the genus 0 case of Theorem~\ref{th:cohft}.  The idea is to produce a 
prepotential from the primary terms of $N_{0,s}(P_1,\dots,P_s)$, i.e., from their constant terms, which uniquely (and constructively) determines a genus 0 CohFT.  Moreover, quasihomogeneity of the prepotential produces a homogeneous CohFT.  The higher coefficients of $N_{0,s}(P_1,...,P_s)$ satisfy a homogeneity condition that forces them to be the correlation functions of the homogeneous CohFT.

The prepotential
\begin{equation}  \label{prep}
F_0=\sum\frac{1}{s!}Y_{0,s}=\frac{1}{2}t_0^2t_1+\sum_{s\geq 3}\frac{1}{s!}N_{0,s}(\vec{0})t_1^s
=\frac{1}{2}t_0^2t_1+\frac{1}{2}(1+t_1)^2\log(1+t_1)-\frac{1}{2}t_1-\frac{3}{4}t_1^2
\end{equation}
assembled from $N_{0,s}(\vec{0})=(-1)^{s-3}(s-3)!$ is quasihomogeneous with respect to the Euler vector field $E=t_0\frac{\partial}{\partial t_0}+2(1+t_1)\frac{\partial}{\partial t_1}$:
$$E\cdot F_0=4F_0+t_1^2+t_0^2.
$$
This ensures that the genus 0 CohFT $I_{0,s}$ produced from Theorem~\ref{th:manin} satisfies
\begin{equation}  \label{pushf}
\pi_*I_{g,s+1}(e_S\otimes e_1)=\frac{1}{2}(1-g+s-\deg-\sum\alpha_{i_k})I_{g,s}(e_S)
\end{equation}
where $e_S=e_{i_1}\otimes...\otimes e_{i_s}$, and $\alpha_0=1$, $\alpha_1=2$ are the coefficients of $E$.  

The CohFT also satisfies the pull-back condition described above
$$
I_{g,s+1}(v_1\otimes...\otimes v_s\otimes e_0)=\pi^*I_{g,s}(v_1\otimes...\otimes v_s).
$$
It is proved in genus 0 via construction, using the fact that the only non-zero genus 0 primary invariant with $e_0$ as an input is $I_{0,3}(e_0\otimes e_0\otimes e_1)=1$.

Teleman \cite{TelStr} produces a unique homogeneous CohFT for all genus extending the homogeneous genus 0 theory.

\begin{theorem}[Teleman \cite{TelStr}] 
A semi-simple homogenous CohFT with flat identity is uniquely and explicitly reconstructible from genus zero data. 
\end{theorem}
Thus, given the genus 0 primary invariants $N_{0,s}(\vec{0})$ there is a unique homogenous CohFT with flat identity.  Below we will see that its correlators agree with the coefficients of $N_{g,s}(P_1,\dots,P_s)$.

The pushforward relation (\ref{pushf}) expressed in terms of correlators is 
\begin{eqnarray*}  \label{pushfcor}
&{}&\int_{\overline{\modm}_{g,s+1}}\hspace{-3mm}I_{g,s+1}(e_S\otimes e_1)\prod_{i=1}^s\psi_i^{k_i}
=\int_{\overline{\modm}_{g,s}}\hspace{-3mm}\pi_*\Big( \prod_1^s(\pi^*\psi_i)^{k_i}+\sum_jD_j\cdot\prod\pi^*\psi_i^{k_i-\delta_{ij}}\Big)I_{g,s+1}(e_S\otimes e_1)\nonumber\\
&{}&\qquad\qquad=\int_{\overline{\modm}_{g,s}}\prod_1^s\psi_i^{k_i}\pi_*I_{g,s+1}(e_S\otimes e_1)+\sum_j
\prod_1^s\psi_i^{k_i-\delta_{ij}}\pi_*(D_j\cdot I_{g,s+1})(e_S\otimes e_1)\\
&{}&\qquad\qquad=\Bigl(\frac{1}{2}\sum_{i=1}^s k_i+\chi_{g,s}\Bigr)
\int_{\overline{\modm}_{g,s}}I_{g,s}(e_S)\prod_{i=1}^s\psi_i^{k_i}+\sum_{j=1}^s\int_{\overline{\modm}_{g,s}}I_{g,s}(e_{S\backslash\{j\}}\otimes e^*_j)\prod_{i=1}^s\psi_i^{k_i-\delta_{ij}}\nonumber
\end{eqnarray*}
which uses
$$\psi_i=\pi^*\psi_i+D_i\Rightarrow \psi_i^k=(\pi^*\psi_i)^k+D_i\cdot\sum\psi_i^m(\pi^*\psi_i)^{k-1-m}
$$
and 
$$D_j\cdot I_{g,s+1}(e_S\otimes e_1)=I_{g,s}\otimes I_{0,3}(e_{S\backslash\{j\}}\otimes\Delta\otimes e_{i_j}\otimes e_1)=I_{g,s}(e_{S\backslash\{j\}}\otimes e^*_{i_j})
$$
for $e^*_i=e_{1-i}$.

The condition $E\cdot F_0=4F_0+t_1^2+t_0^2$ on $N_{0,s}(\vec{0})$ is a specialisation to $g=0$ and $P_i=0$ of the divisor equation \cite{NorStr}
\begin{equation}   \label{divisor}
N_{g,s+1}(0,P_1,...,P_s)=\sum_{j=1}^s\sum_{k=1}^{P_j-1}kN_{g,s}(P_1,...,P_s)|_{P_j=k}+\left(\frac{1}{2}\sum_{j=1}^s P_j+\chi_{g,s}\right)N_{g,s}(P_1,\dots,P_s)
\end{equation} 
which is exactly the pushforward relation on correlators (\ref{pushfcor}).

The flat identity pull-back condition is known as the {\em string equation} on correlators for $2g-2+s>0$:
$$\int_{\overline{\modm}_{g,s+1}}I_{g,s+1}(v_1\otimes\cdots\otimes v_s\otimes e_0)\prod_{i=1}^s\psi_i^{k_i}
=\sum_{j=1}^s\int_{\overline{\modm}_{g,s}}I_{g,s}(v_1\otimes\cdots\otimes v_s)\prod_{i=1}^s\psi_i^{k_i-\delta_{i,j}}
$$
and agrees with the recursion \cite{NorStr}
\begin{equation}   \label{string}
N_{g,s+1}(1,P_1,\dots,P_s)=\sum_{j=1}^s\sum_{k=1}^{P_j}kN_{g,s}(P_1,\dots,P_s)|_{P_j=k}
\end{equation}
In particular, this proves the genus 0 case of Theorem~\ref{th:cohft} since the recursions (\ref{divisor}) and (\ref{string}) uniquely determine the correlators of $I_{0,s}$ and $N_{0,s}(P_1,\dots,P_s)$.

\subsection{Explicit description of CohFT in genus 0}

This constructive proof gives rise to a rather explicit description of the genus 0 classes $I_{0,s}(e_S)\in H^*(\overline{\modm}_{0,s})$.  It is sufficient to describe the pairing of the class $I_{0,s}(e_S)$ with all strata $\overline{D}_{\Gamma}$ since this uniquely characterises $I_{0,s}(e_S)$. 
\begin{proposition} We have that
\begin{equation}  \label{genus0}
\int_{\overline{D}_{\Gamma}} I_{0,s}(e_S)=\left\{\begin{array}{c}\chi(D_{\Gamma})\\ 0\end{array}\right.
\end{equation}
where evaluation of $\chi(D_{\Gamma})$ or 0 can be characterised explicitly in terms of $\Gamma$ and $e_S=e_{i_1}\otimes...\otimes e_{i_s}$.
\end{proposition}
{\bf Proof.}\ 
If $\Gamma$ consists of a single vertex, so that $\overline{D}_{\Gamma}=\overline{\modm}_{0,s}$ then (\ref{genus0}) holds by construction---it is the constant term of $N_{0,s}(P_1,\dots,P_s)$.  In other words
$$\int_{\overline{\modm}_{0,s}} I_{0,s}(e_S)=\left\{\begin{array}{cc}\chi(\modm_{0,s})&e_S=e_1^{\otimes s}\\ 0&\text{otherwise}\end{array}\right..
$$

Given a dual graph $\Gamma$ of type $(0,s)$, its associated strata is a product of (open) moduli spaces, one  factor for each vertex of $\Gamma$, 
$$D_{\Gamma}=\modm_{0,s_1}\times...\times \modm_{0,s_{|V(\Gamma)|}}$$ 
where $s_j$ is the valence of the $j$th vertex of $\Gamma$.  Hence
$$\chi(D_{\Gamma})=\chi(\modm_{0,s_1})\cdot...\cdot \chi(\modm_{0,s_k}).$$ 
In terms of the primary correlators $Y_{0,s_i}=\int_{\overline{\modm}_{0,s_i}}I_{0,s_i}:H^{\otimes s_i}\to\bc$ 
$$ \int_{\overline{D}_{\Gamma}}I_{0,s}(e_S)=\bigotimes_{v\in V_{\Gamma}}Y_{0,|v|}\left(e_S\otimes\Delta^{\otimes |E_{\Gamma}|}\right)=c(\Gamma,e_S)\cdot\chi(\modm_{0,s_1})\cdot...\cdot \chi(\modm_{0,s_k})=c(\Gamma,e_S)\cdot\chi(D_{\Gamma})
$$
for some $c(\Gamma,e_S)\in\bz$ since each primary correlator is either zero or an Euler characteristic.  Given $e_S$, if there is an assignment of $e_0\otimes e_1$ or $e_1\otimes e_0$ to each interior edge which gives a non-zero evaluation, then it is unique since $\Gamma$ is a tree.  Hence $c(\Gamma,e_S)=1$ or 0 as required.\ \ $\Box$

\begin{remark}
More generally the same argument shows that for any genus
$$
\int_{\overline{D}_{\Gamma}} I_{g,s}(e_S)=c(\Gamma,e_S)\dot\chi(D_{\Gamma})
$$
for $c(\Gamma,e_S)\in\bz^{\geq 0}$.  The integer $c(\Gamma,e_S)$ can take values other than 0 or 1 since there can be more than one assignment of $e_0\otimes e_1$ or $e_1\otimes e_0$ to each interior edge which gives a non-zero evaluation.
\end{remark}

\subsection{General proof of Theorem~\ref{th:cohft} using DOSS method \cite{DOSSIde}.}

In higher genus we have agreement of relations among the correlators of the CohFT evaluated on boundary classes and relations among the coefficients of $N_{g,s}(P_1,\dots,P_s)$, but this is not quite enough to prove that they coincide.  Instead we apply the results of \cite{DOSSIde} where it is shown that if the spectral curve satisfies a constraint then the Givental reconstruction of higher genus correlators can be understood in terms of graphs, and the same graphs are used to calculate topological recursion.  

Dunin-Barkowsky, Orantin, Shadrin and Spitz \cite{DOSSIde} using Eynard's technique of \cite{Ey11}
associated to any semi-simple CohFT a local spectral curve $(\Sigma,B,x,y)$.  The $R$-matrix gives rise to the bidifferential $B$ on the spectral curve 
\be\label{Givental1}
\check{B}^{i,j}_{p,q}=[z^pw^q]\frac{\delta^{ij}-\sum_{k=1}^NR^i_k(-z)R^j_k(-w)}{z+w}
\ee
where $\check{B}^{i,j}_{p,q}$ are coefficients of an asymptotic expansion of the Laplace transform of the regular part of $B$ expressed in terms of the local coordinates $s_i=\sqrt{x-x(a_i)}$ where $dx(a_i)=0$.  See \cite{DOSSIde} for details.  The $R$-matrix together with the transition matrix $\Psi$ from a flat to a normalised canonical bases gives rise to the meromorphic differential $ydx$ in terms of $s_i$.  In particular, this implies a compatibility condition between the differential $ydx$ and the bifferential $B$.  It is given by (\ref{compatible}) below.  Most spectral curves will fail this condition.


One can apply \cite{DOSSIde} in either direction, beginning with a semi-simple CohFT or a spectral curve.  The prepotential $F_0$ (\ref{prep}) gives rise to a semi-simple CohFT.  It can be used to produce the $R$-matrix and transition matrix $\Psi$ and hence the spectral curve.  Instead, since we already have a candidate for the spectral curve we will start with the spectral curve and apply \cite{DOSSIde} to get the coefficients of $N_{g,s}(P_1,\dots,P_s)$ as ancestor invariants of a CohFT.  Since it agrees with the CohFT above in genus 0, by uniqueness it is the same CohFT produced by Teleman's theorem.

The spectral curves for the \dvol\  and Gromov-Witten invariants of $\bp^1$ are similar:
$$x=z+1/z,\quad y=z,\quad B=\frac{dzdz'}{(z-z')^2}$$
$$x=z+1/z,\quad y=\ln{z},\quad B=\frac{dzdz'}{(z-z')^2}$$
and since $x$ and $B$ determine the $R$-matrix, it is the same for both curves.  The $R$-matrix for Gromov-Witten invariants of $\bp^1$ was calculated explicitly in \cite{DOSSIde}:
$$
R(u)=\sum_{k=0}^{\infty}R_ku^k,\quad R_k=\frac{(2k-1)!!(2k-3)!!}{2^{4k}k!}
\left(\begin{array}{cc}-1&(-1)^{k+1}2ki\\2ki&(-1)^{k+1}\end{array}\right).
$$
The results of \cite{DOSSIde} can be applied to spectral curves such that a Laplace transform of $ydx$ is related to this $R$-matrix (which is essentially the Laplace transform of the regular part of the bidifferential).

For local coordinates $s_i$, $i=1,2$ near $x=\pm 2$ given by $x=s_i^2\pm 2$ 
$$ y= 1+s_1+\frac{1}{2}s_1^2+\sum_{k=1}^{\infty}(-1)^{k-1}\frac{(2k-3)!!}{2^{3k}k!}s_1^{2k+1}$$
$$ y= -1+is_2+\frac{1}{2}s_2^2-i\sum_{k=1}^{\infty}\frac{(2k-3)!!}{2^{3k}k!}s_2^{2k+1}$$
$$\check{(ydx)}_1=\frac{\sqrt{u}}{2\sqrt{\pi}}\int_{\gamma_1} e^{-u(x-2)}ydx\sim\sum_{k=0}^{\infty}(-1)^{k-1}\frac{(2k+1)!!(2k-3)!!}{2^{4k+1}k!}u^{-(k+1)}
$$
$$\check{(ydx)}_2=\frac{\sqrt{u}}{2\sqrt{\pi}}\int_{\gamma_2} e^{-u(x+2)}ydx\sim-i\sum_{k=0}^{\infty}\frac{(2k+1)!!(2k-3)!!}{2^{4k+1}k!}u^{-(k+1)}
$$
We use the convention $(-1)!!=1$, $(-3)!!=-1$ and $\sim$ means Poincare asymptotic in the parameter $u$.

The compatibility condition between the differential $ydx$ and the bifferential (appearing here in terms of the $R$-matrix which is related to the bidifferential via the Laplace transform) is
\begin{equation}  \label{compatible}
\frac{1}{\sqrt{2}}\left(\begin{array}{cc}1&i\end{array}\right)\cdot\frac{1}{\sqrt{2}}R(u)=\left(\begin{array}{cc}\check{(ydx)}_1&\check{(ydx)}_2\end{array}\right)
\end{equation}
which uses the first row of the transition matrix
$$
\Psi=\frac{1}{\sqrt{2}}\left(\begin{array}{cc}1&i\\1&-i\end{array}\right).
$$
One can check directly that it is satisfied for $x=z+1/z$, $y=z$, $B=dzdz'/(z-z')^2$.

From this, \cite{DOSSIde} supplies the times
$$
\xi^0_0=\frac{1}{2}\left(\frac{1}{1-z}-\frac{1}{1+z}\right)
$$
$$
\xi^1_0=\frac{1}{2}\left(\frac{1}{1-z}+\frac{1}{1+z}\right)
$$
\be
\label{question}
\xi^i_k=\left(\frac{d}{dx}\right)^k\xi^i_0=\sum_k p_{k,i}z^k
\ee
and the main result
$$W^{(g)}_s(x_1,...,x_s)=\sum_{\vec{k},\vec{\alpha}} c^g_{\vec{k},\vec{\alpha}} \prod_{i=1}^n \xi_{k_i,\alpha_i}$$
where the coefficients are the ancestor invariants (\ref{ancestor}).
As remarked above, the CohFT produced this way necessarily coincides with the homogeneous CohFT produced by Teleman's theorem since they both use Givental reconstruction and the same initial data.

\subsection{Euler characteristic of the moduli space of curves}

The primary correlators of the CohFT described above are given by the orbifold Euler characteristic $\chi(\modm_{g,s})$.  This is perhaps not so deep in genus 0 where it is simply the statement that
$$\int_{\overline{\modm}_{0,s}}I_{0,s}(e_1^{\otimes s})=\left.\frac{\partial^sF}{\partial t_{1}^s}\right|_{(t_0,t_1)=(0,1)}\hspace{-8mm}=\hspace{4mm}\chi(\modm_{0,s})$$
for $F=\frac{1}{2}t_{0,0}^2t_{0,1}+\frac{1}{2}t_{0,1}^2\log{t_{0,1}}$ given in (\ref{Frob}).
It is a deeper property that
$$\int_{\overline{\modm}_{g,s}}I_{g,s}(e_1^{\otimes s})=\chi(\modm_{g,s})$$
and it is the purpose of this section to show that homogeneity of the CohFT goes some way towards explaining the appearance of $\chi(\modm_{g,s})$.

The exact sequence of mapping class groups
$$
1\rightarrow\pi_1(C-\{p_1,...,p_s\})\rightarrow\Gamma_g^{s+1}\rightarrow\Gamma_g^s\rightarrow 1
$$
implies $\chi(\Gamma_g^s)=\chi(\Gamma_g^{s+1})/\chi(C-\{p_1,...,p_s\})$.  Since the (orbifold) Euler characteristic is $\chi(\modm_{g,s})=\chi(\Gamma_g^s)$ then
\begin{equation}  \label{eulrec}
\chi(\modm_{g,s+1})=(2-2g-s)\cdot\chi(\modm_{g,s})\quad\text{for}\quad2g-2+s>0
\end{equation} 
and $\chi(\modm_{g,1})=\zeta(1-2g)$ \cite{HZ,PenPer}.

The relation (\ref{eulrec}) arises naturally out of the push-forward relation satisfied by the CohFT which restricts to the top degree terms.  
\begin{equation}    \label{cohfthom}
\pi_*I_{g,s+1}(e_{i_1}\otimes...\otimes e_{i_s}\otimes e_1)=\frac{1}{2}\left(1-g+s-\deg-\sum\alpha_{i_k}\right)I_{g,s}(e_{i_1}\otimes\cdots\otimes e_{i_s})
\end{equation}
Note that it is usually more natural to consider the degree 0 part of the CohFT---a topological field theory--- since the pull-back relations satisfied by a CohFT restrict to degree 0.

Denote the degree $3g-3+s$ part of $I_{g,s}(e_1^{\otimes s})$ by $c_{g,s}$ so
$$I_{g,s}(e_1^{\otimes s})=\epsilon(g+s)2^g+...+c_{g,s}\{\text{pt}\}$$
where $\epsilon(g+s)\in\{0,1\}$ is the mod 2 reduction of $g+s$.  Note that all insertions are necessarily $e_1$ since any $e_0$ insertion comes from a pull-back
$$\deg I_{g,s+1}(e_0\otimes e_S)=\deg \pi^*I_{g,s}(e_S)=\deg \pi^*I_{g,s}(e_S)<3g-3+s+1$$
and hence is not of top degree.  The top degree class is well-behaved under push-forward.
$$ \pi_*c_{g,s+1}\{\text{pt}\}=\frac{1}{2}\left(1-g+s-(3g-3+s)-2s\right)c_{g,s}\{\text{pt}\}=(2-2g-s)c_{g,s}\{\text{pt}\}
$$
hence
$$c_{g,s+1}=(2-2g-s)c_{g,s}\Rightarrow c_{g,s}=c_g\cdot \chi(\modm_{g,s}).$$
This explains the appearance of the Euler characteristic $\chi(\modm_{g,s})$, up to a factor, in this CohFT.  It turns out that $c_g=1$ but this needs the \dvol\  to prove.

\subsection{Ancestor invariants and Gaussian means}\label{ss:Ancestor-Gauss}

From Lemma~\ref{lm:Norbury} and the formulae (\ref{W1-1}) and (\ref{W1-as}), we immediately obtain
formula for the loop means in terms of  the ancestor invariants.

\begin{theorem}\label{th:Wgs2ckl}
We have the following explicit relation between the ancestor invariants (\ref{ancestor}) of a CohFT
and the Gaussian means:
\beq
W^{(g)}_s\bigl(e^{\lambda_1}+e^{-\lambda_1},\dots,e^{\lambda_s}+e^{-\lambda_s}\bigr)
=\sum_{\vec{k},\vec{\alpha}}c^g_{\vec{k},\vec{\alpha}}\prod_{j=1}^s \widehat p_{k_j,\alpha_j}(\lambda_j),
\eeq
where
\beq
\widehat p_{k,\alpha}(\lambda)=\left\{\begin{array}{lll}
2^{1-2r}(2r+1)s_{r,0}(\lambda),&k=2r,&\alpha=0;\cr
2^{-2r}(2r+1)s_{r,1}(\lambda),&k=2r,&\alpha=1;\cr
2^{-2r}(2r+2)(2r+3)s_{r+1,1}(\lambda),&k=2r+1,&\alpha=0;\cr
2^{-2r-1}s_{r,0}(\lambda),&k=2r+1,&\alpha=1,
\end{array}
\right.
\label{pk}
\eeq
and $s_{r,\beta}(\lambda)$, $\beta=0,1$, are defined in (\ref{sj}).
\end{theorem}

\begin{example}\label{ex:top}
The topological (= degree zero) part of the CohFT is
$$I_{g,s}(e_{\alpha_1}\otimes\cdots\otimes e_{\alpha_s})=\epsilon(\vec{\alpha})2^g+\ \hbox{higher degree terms}$$
where $\displaystyle \epsilon(\vec{\alpha})\equiv\sum_{i=1}^s\alpha_i$ (mod 2) is 0 or 1.
This explains the asymptotic behaviour of the topological invariants $W^{(g)}_{s}$ at their poles.
\end{example}
\begin{example}\label{ex:volume}
If $\{e_0,e_1\}$ is a basis of $H$ corresponding to flat coordinates then
$$\int_{\overline{\modm}_{g,s}}I_{g,s}(e_1^{\otimes s})=\chi(\modm_{g,s}).$$
This uses the fact that $N_{g,s}(0,0,\dots,0)=\chi(\modm_{g,s})$ and
$$p_{k,\alpha}(0)=\left\{\begin{array}{cc} 1,&(k,\alpha)=(0,0)\\
 0,& \text{otherwise}.\end{array}\right.
 $$
\end{example}

{\sl We thus identify the coefficients $\widehat{b}^g_{\vec{k},\vec{\beta}}$ of the expansions (\ref{W->sk}) with (linear 
combinations) of the ancestor invariants $c^g_{\vec{k},\vec{\alpha}}$ using the identification} (\ref{pk}): for $s=1$, we have
\bea
&{}&\widehat b^g_{r,0}=2^{1-2r}(2r+1)c^g_{2r,0}+2^{-1-2r}c^g_{2r+1,1},\nonumber\\
&{}&\widehat b^g_{r,1}=2^{-2r}(2r+1)c^g_{2r,1}+2^{2-2r}2r(2r+1)c^g_{2r-1,0}.\nonumber
\eea

\section{The one-backbone case. The Harer--Zagier formula}
\label{s:HZ}

\subsection{Times in the one-backbone case.}

The formulas (\ref{main}) and (\ref{times}) become especially useful in the one-backbone case. Then,
we have just one variable $\lambda$ and just one set of times, and we always have the sum (\ref{time-w})
in the r.h.s. of (\ref{main}). It is useful to introduce the special notation for the one-loop resolvent
\beq
L_g(x):=W_1^{(g)}(x).
\eeq
We have just one external edge in this case, so from (\ref{main}) we merely have
\beq
L_g(e^{\lambda}+e^{-\lambda})=\sum_{r=0}^{3g-2}
(-1)^r \frac{\varkappa_{g,1,r}}{2^{d-r}(d-r)!}\frac{1}{e^{\lambda}-e^{-\lambda}}
\left(\frac{\partial}{\partial \lambda}\right)^{2d-2r+1}\frac{2}{e^{2\lambda}-1},\quad d=3g-2,
\label{one-bone}
\eeq
where $\varkappa_{g,1,r}$ are (conjecturally positive) rational numbers, $\varkappa_{g,1,0}=\<\tau_{3g-2}\>_g$.

\subsection{The recurrence relations}

The expansion coefficients $c_g(k)$ (in the original notation) of the loop mean
\beq
\label{Lgw}
L_g(w)=
\sum_{k=2g}^\infty w^{-2k-1}\left\langle\tr H^{2k}\right\rangle_g=\sum_{k=2g}^\infty c_g(k)w^{-2k-1}
\eeq
satisfy the celebrated Harer--Zagier three-term recursion relation~\cite{HZ}:
\beq
\label{HZ}
(n+1)c_g(n)=2(2n-1)c_g(n-1)+(2n-1)(n-1)(2n-3)c_{g-1}(n-2).
\eeq

\begin{lemma}
\label{lem-rec}
The generating function $L_g(w)$ satisfy the differential recurrence relation
\beq
L'''_g(w)=(w^2-4)L'_{g+1}(w)-wL_{g+1}(w),
\eeq
which, upon substitution $w=e^{\lambda}+e^{-\lambda}$, takes the form
\beq
\label{L-lambda}
L'''_g(e^{\lambda}+e^{-\lambda})=(e^{\lambda}-e^{-\lambda})^2L'_{g+1}(e^{\lambda}+e^{-\lambda})
-(e^{\lambda}+e^{-\lambda})L_{g+1}(e^{\lambda}+e^{-\lambda}).
\eeq
\end{lemma}

Note the appearance of the factor $(e^{\lambda}-e^{-\lambda})^2$ of the first term in the r.h.s. It simplifies
the further recurrence procedure drastically.

The next proposition is a corollary of Lemma~\ref{lm:multiloop}.

\begin{proposition}
\label{lm1}
The general form of $L_g(e^{\lambda}+e^{-\lambda})$ is (cf. Example~\ref{ex:Ms1})
\beq
\label{general}
L_g(e^{\lambda}+e^{-\lambda})
=\frac{1}{(e^\lambda-e^{-\lambda})^{4g+1}}\sum_{k=0}^{g-1}\frac{b^{(g)}_k}{(e^{\lambda}-e^{-\lambda})^{2k}}.
\eeq
\end{proposition}



\begin{remark}\label{rm1}
Note that the coefficients $b^{(g)}_k$ are the coefficients
$\kappa_g(n)$ appeared in the original paper~\cite{HZ} by Harer and Zagier. But the three-term recurrence relation on
them was not written there. We present it below. This relation was also found in \cite{HT}.
\end{remark}

\begin{proposition}
\label{th5}
The coefficients $b^{(g)}_k$ in Proposition~\ref{lm1} satisfy the three-term recurrence relation:
\beq
\label{recurrence}
(4g+2k+6)b^{(g+1)}_k=(4g+2k+1)(4g+2k+3)\Bigl[(4g+2k+2)b^{(g)}_k+4(4g+2k-1)b^{(g)}_{k-1}\Bigr].
\eeq
\end{proposition}

{\em Proof}. The proof is the direct substitution, we just mention the basic steps. We first find the
action of the third derivative on any monomial $(e^{\lambda}-e^{-\lambda})^{-t}$:
$$
\left[\frac{1}{e^{\lambda}-e^{-\lambda}}\frac{d}{d\lambda}\right]^3\frac{1}{(e^{\lambda}-e^{-\lambda})^t}
=-t(t+2)\left[(t+1)\frac{e^{\lambda}+e^{-\lambda}}{(e^{\lambda}-e^{-\lambda})^{t+4}}
+4(t+4)\frac{e^{\lambda}+e^{-\lambda}}{(e^{\lambda}-e^{-\lambda})^{t+6}}\right].
$$
The result thus contains two terms,
whereas the action of the operator in the right-hand side of (\ref{L-lambda}) on the same monomial is merely
$$
\left[(e^{\lambda}-e^{-\lambda})\frac{d}{d\lambda}-(e^{\lambda}+e^{-\lambda})\right]\frac{1}{(e^{\lambda}-e^{-\lambda})^t}
=-(t+1)\frac{e^{\lambda}+e^{-\lambda}}{(e^{\lambda}-e^{-\lambda})^{t}}
$$
and it contains just one term. Substituting expansions (\ref{general}) in both sides of (\ref{L-lambda}) and equating
coefficients at equal powers of $\frac{e^{\lambda}+e^{-\lambda}}{(e^{\lambda}-e^{-\lambda})^{s}}$ we obtain (\ref{recurrence}),
which completes the proof.

\begin{lemma}\label{lm2}
All the coefficients $b^{(g)}_k$ are positive integers.
\end{lemma}

{\em Proof}. The positivity obviously follows from the fact that the both coefficients are positive in the r.h.s. of the recurrence relation
(\ref{recurrence}). The integrality follows from the comparison with the  expansion of the original Harer--Zagier function
$L_g(e^\lambda+e^{-\lambda})$ at large positive $\lambda$. Indeed, comparing formulas
(\ref{Lgw}) and (\ref{general}) and expanding in $e^{-\lambda}$ it is straightforward to obtain that
\bea
L_g(e^\lambda+e^{-\lambda})&=&\sum_{k=2g}^\infty\sum_{p=0}^\infty c_g(k)(-1)^p\left({p\atop 2k+p}\right)e^{-(2k+2p+1)\lambda}
\nonumber\\
&=&\sum_{k=0}^{g-1}\sum_{p=0}^\infty b^{(g)}_k \left({p\atop 4g+2k+p}\right)e^{-(4g+2k+2p+1)\lambda}.
\label{comparison}
\eea
All the Harer--Zagier coefficients $c_g(k)$ are positive integers. If we use formula (\ref{comparison}) to express
$b^{(g)}_k$ through $c_g(k)$ equating the terms at equal powers of $e^{-\lambda}$, we obtain for the coefficient
of $e^{-(4g+1)\lambda}$ the equality
$$
b^{(g)}_0=c_g(2g)
$$
and $b^{(g)}_0$ is therefore an integer. The equality at $e^{-(4g+3)\lambda}$ reads
$$
b^{(g)}_1+\left(1\atop 4g+1\right)b^{(g)}_0=c_g(2g+1)-\left(1\atop 4g+1\right)c_g(2g),
$$
so $b^{(g)}_1$ is an integer as well, etc. Every first appearance of any coefficient $b^{(g)}_k$ in this chain of recurrence
equations is with the unit coefficient,
which enables us to express $b^{(g)}_k$ as a polynomial in the preceding $b^{(g)}_p$, $p=0,\dots,k-1$, and
in $c_g(2g+q)$, $q=0,\dots,k$, with integer coefficients, so, obviously, $b^{(g)}_k$ is an integer, which completes the
proof.

\begin{conjecture}\label{con:integrality}
Because we have identified the coefficients $\widehat b^{(g)}_{\vec k,\vec \beta}$ with combinations of the ancestor
invariants in the general (multi-backbone) case as well, we put forward the conjecture that all these coefficients are
positive integers.
\end{conjecture}


In order to simplify expressions it is convenient to introduce the {\bf renormalized} $b^{(g)}_k$ by the formula
\beq
\label{ren-b}
{b}^{(g)}_k = {\mathfrak b}^{(g)}_k(6g-1-2k)!!.
\eeq
In terms of these renormalized quantities, the general recursion relation (\ref{recurrence}) 
looks especially simple:
\beq
\label{recurrence-renorm}
(2g+k+3){\mathfrak b}^{(g+1)}_k=(2g+k+1){\mathfrak b}^{(g)}_k+2{\mathfrak b}^{(g)}_{k-1},\quad
{\mathfrak b}^{(1)}_0=\frac13.
\eeq

\subsection{Examples}

Note first that $b_0^{(1)}=1$, which follows from the simple diagrammatic calculation in the genus-one case.
We obtain the coefficients $b_0^{(g)}$ and $b_{g-1}^{(g)}$ from the simple one-term recurrence relations
obtained by setting respectively $k=0$ and $k=g$ in (\ref{recurrence}):
\bea
(4g+6)b_0^{(g+1)}&=&(4g-1)(4g+3)(4g+2)b_0^{(g)},\\
(6g+6)b_g^{(g+1)}&=&4(6g+1)(6g+3)(6g-1)b_{g-1}^{(g)},
\eea
which immediately give
\beq
\label{b0-bg}
b_{g-1}^{(g)}=\frac{2^{g-1}\,(6g-3)!!}{3^g\, g!},\qquad b_{0}^{(g)}=\frac{(4g)!}{8^g\,g!\,(2g+1)!!}.
\eeq
Note that in terms of the renormalized quantities (\ref{ren-b}) we merely have
\beq
\label{b0-bg-ren}
{\mathfrak b}^{(g)}_{g-1}=\frac{2^{g-1}}{3^g\, g!},\qquad {\mathfrak b}^{(g)}_{0}=\frac{1}{2g+1}.
\eeq

Now substitute
$b_{g-1}^{(g)}$ into (\ref{one-bone}) and evaluate the term without reduction ($r=0$) which contains the
highest Kontsevich coefficient $\left\langle \tau_{3g-2}\right\rangle_g$. The easiest way to
do it is to compare the highest-order poles in the both expressions. The result reads
\beq
\left\langle \tau_{3g-2}\right\rangle_g=\frac{1}{2^{3g}\,3^g\,g!}
\eeq
and it is in the complete agreement with the standard formulas.

\subsection{Solving recursion equations for $b^{(g)}_{g-1-k}$ at fixed $k$ and arbitrary $g$}\label{ss:bg-1-k}

\subsubsection{The coefficient $b^{(g)}_{g-2}$}\label{sss:bg-2-HZ}

The recurrence relation for the renormalized ${\mathfrak b}^{(g)}_{g-2}$ reads:
\beq
(3g+2){\mathfrak b}^{(g+1)}_{g-1}=(3g){\mathfrak b}^{(g)}_{g-1}+2{\mathfrak b}^{(g)}_{g-2},\qquad
{\mathfrak b}^{(1)}_{-1}=0.
\label{b11}
\eeq
We try the anzatz ${\mathfrak b}^{(g)}_{g-2}=f(g){\mathfrak b}^{(g)}_{g-1}$. Then, since
${\mathfrak b}^{(g+1)}_{g}=2{\mathfrak b}^{(g)}_{g-1}/(3g-3)$, we obtain for $f(g)$ the relation
$$
(3g+2)f(g+1)\frac{2}{3g+3}=3g+2f(g),
$$
which can be solved by a polynomial substitution provided we can cancel $g+1$ in the denominator in the
right-hand side. So, taking $f(g)=a g(g-1)$, we obtain merely that $(3g+2)(2a/3)=3+2a(g-1)$. For the term linear
in $g$ we have the identity, whereas the constant term gives $a=9/10$. The normalization condition at $g=1$ in
(\ref{b11}) is satisfied, and we therefore obtain a unique solution to the recursion 
relation (\ref{b11}):
\beq
{\mathfrak b}^{(g)}_{g-2}=\frac{9}{10}g(g-1){\mathfrak b}^{(g)}_{g-1},\quad\hbox{i.e.,}\
{\mathfrak b}^{(g)}_{g-2}=\frac15\frac{2^{g-2}}{3^{g-2}(g-2)!},
\quad\hbox{or}\quad
b^{(g)}_{g-2}=\frac 15 \frac{2^{g-2}\,(6g-5)!!}{3^{g-2}\,(g-2)!}.
\label{bgg-2}
\eeq

Expanding in the times $t^{\pm}_{2d+1}(\lambda)$ (see (\ref{one-bone})), for the first ``reduction coefficient''
$\varkappa_{g,1,1}$ we obtain that
\beq
\varkappa_{g,1,1}=\frac 15[12g^2-7g+5]\varkappa_{g,1,0},\quad g\ge 2.
\label{c-g-1-1}
\eeq

\subsubsection{The coefficient $b^{(g)}_{g-3}$}

The recurrence relation (\ref{recurrence}) for $k=g-2$ reads
\beq
\label{k=g-2}
(6g+2)b^{(g+1)}_{g-2}=(6g-1)(6g-3)\bigl[(6g-2)b^{(g)}_{g-2}+4(6g-5)b^{(g)}_{g-3}\bigr].
\eeq
Recall that due to (\ref{bgg-2}) the coefficients $b^{(g)}_{g-2}$ now themselves satisfy the
two-term recursion
\beq
\label{two-term-bg-2}
b^{(g+1)}_{g-1}=\frac{2(6g+1)(6g-1)(6g-3)}{3(g-1)}b^{(g)}_{g-2}
\eeq
and we can try to find an anzatz of form (\ref{bgg-2}) relating now $b^{(g)}_{g-3}$ and $b^{(g)}_{g-2}$.
This anzatz must have a form of a rational function with polynomial of the second order in the numerator. But
it must also cancel the factor $6g-5$ in the recurrence relation. Moreover, the polynomial in the numerator must be
divisible by $g-2$ to cancel the factor $g-1$ in the denominator of the recurrence relation (\ref{two-term-bg-2}).
So, we try the anzatz
\beq\label{anzatz-b-g-3}
\tilde b^{(g)}_{g-3}=\kappa \frac{(g-2)(g-t)}{6g-5}b^{(g)}_{g-2}
\eeq
with unknown $\kappa$ and $t$. Substituting this anzatz into (\ref{k=g-2}), we obtain the simple polynomial equation of second order in $g$:
$$
\frac23(6g+2)\kappa (g-t+1)=(6g-2)+4\kappa(g-2)(g-t)
$$
whose highest term always matches. A unique solution is $t=1/2$, $\kappa=3/10$. But if we substitute this solution into expression (\ref{b0-bg})
for the term $b^{(3)}_0$ we observe the mismatch, which is due to the fact that we are able to add to the obtained solution
$\tilde b^{(g)}_{g-3}$ of the inhomogeneous equation (\ref{k=g-2}) any solution $\hat b^{(g)}_{g-3}$
of the {\em homogeneous} recursion relation
\beq
\label{hat-b-g-3}
(6g+2){\hat b}^{(g+1)}_{g-2}=4(6g-1)(6g-3)(6g-5){\hat b}^{(g)}_{g-3},
\eeq
so, with the proper normalization for the term $b^{(3)}_0$, we obtain $b^{(g)}_{g-3}$ as a {\em sum of two terms}:
\beq
\label{b-g-3}
b^{(g)}_{g-3}=\frac{(2g-1)\,2^{g-3}\,(6g-7)!!}{5^2\,3^{g-3}\,(g-3)!}-\frac{7\,2^{g-3}\,(6g-7)!!}{10\,(3g-2)!!!},
\ \hbox{where}\ (3g-2)!!!\equiv \prod_{k=3}^g(3k-2).
\eeq

The general structure of solution becomes clear: in order to find
$b^{(g)}_{g-4}$ we first solve {\em two} inhomogeneous equations
with ${\tilde b}^{(g)}_{g-3}$ and ${\hat b}^{(g)}_{g-3}$ in the
right-hand sides and then again add the solution of the
homogeneous equation with the proper coefficient to satisfy the
initial condition for $b^{(g)}_0$. We now formulate the {\em
general} procedure for finding all coefficients $b^{(g)}_{g-1-k}$ at
all $g$ and fixed $k$.

\subsubsection{Finding $b^{(g)}_{g-1-k}$}

We formulate the general procedure in terms of the renormalized
coefficients ${\mathfrak b}^{(g)}_{g-1-k}$ (\ref{ren-b}). The
procedure consists of several steps.

{\bf Step 1.} Solving the homogeneous equation for ${\hat{\mathfrak b}}^{(g)}_{g-1-s}$ enumerated by $s=0,1,\dots$:
\beq
\label{rec-s}
(3g-s+3){\wht{\mathfrak b}}^{(g+1)}_{g-s}=2{\wht{\mathfrak b}}^{(g)}_{g-s-1},\ \hbox{or}\
{\wht{\mathfrak b}}^{(g+1)}_{g-s}=\frac{1}{q(g+1,s)}{\wht{\mathfrak b}}^{(g)}_{g-s-1},
\eeq
where
\beq
\label{q-fun}
q(g,s):=\frac12 (3g-s), \quad s=0,1,\dots.
\eeq
Then, obviously,
\beq
\label{b-hat}
{\wht{\mathfrak b}}^{(g)}_{g-s-1}=\Bigl[\prod_{l=1}^{g-s-1}q(s+1+l,s)\Bigr]^{-1}{\wht{\mathfrak b}}^{(s+1)}_{0}
\eeq

{\bf Step 2.} Solving the chain of inhomogeneous equations for ${\wtd{\mathfrak b}}^{(g)}_{g-s-1-k}$ that are solutions of the equations
\beq
q(g+1,s+k){\wtd{\mathfrak b}}^{(g+1)}_{g-s-k}=q(g,s+k-1){\wtd{\mathfrak b}}^{(g)}_{g-s-k}+
{\wtd{\mathfrak b}}^{(g)}_{g-s-k-1}
\eeq
by substituting the {\em anzatz}
\beq
\label{anzatz-general}
{\wtd{\mathfrak b}}^{(g)}_{g-s-k-1}=P_{(s,k)}(g)q(g-k+1,s)
\left[\frac{{\wtd{\mathfrak b}}^{(g)}_{g-s-k}}{P_{(s,k-1)}(g)}\right],
\eeq
where
\beq
{\wtd{\mathfrak b}}^{(g)}_{g-s-1}={\wht{\mathfrak b}}^{(g)}_{g-s-1}
\eeq
and all $P_{(s,k)}(g)$ are polynomials of order $k$ to be determined ($P_{(s,0)}(g)=1$).

It is easy to see, by recursion, that the quantities
$\left[\frac{{\wtd{\mathfrak b}}^{(g)}_{g-s-k}}{P_{(s,k-1)}(g)}\right]$ satisfy the recursion relation
$$
\left[\frac{{\wtd{\mathfrak b}}^{(g+1)}_{g+1-s-k}}{P_{(s,k-1)}(g+1)}\right]
=q^{-1}(g+2-k,s)\left[\frac{{\wtd{\mathfrak b}}^{(g)}_{g-s-k}}{P_{(s,k-1)}(g)}\right],
$$
whence, for $P_{(s,k)}(g)$, we have the recursion
\beq
\label{rec-P}
(3g+3-k-s)P_{(s,k)}(g+1)=(3g+1-k-s)P_{(s,k-1)}(g)+2q(g-k+1,s)P_{(s,k)}(g),\quad P_{(s,0)}=1,
\eeq
or, explicitly,
\beq
\label{rec-P-exp}
(3g+3-k-s)P_{(s,k)}(g+1)=(3g+1-k-s)P_{(s,k-1)}(g)+(3g-3k+3-s)P_{(s,k)}(g),\quad P_{(s,0)}=1.
\eeq
Solving this relation, we can find all the polynomials $P_{(s,k)}(g)$ starting from the zeroth one.
This recursion always has a solution because
every highest term relation (the coefficient of $g^{k+1}$)
is satisfied automatically and at each step we have a system of $k+1$ inhomogeneous linear equations
on $k+1$ coefficients of the polynomial $P_{(s,k)}(g)$.

{\bf Step 3.} In light of 
(\ref{b-hat}), we now express ${\wtd{\mathfrak b}}^{(g)}_{g-s-k-1}$ as 
\beq
{\wtd{\mathfrak b}}^{(g)}_{g-s-k-1}=P_{(s,k)}(g)\prod_{l=1}^{g-s-1-k}q(s+1+l,s){\wht{\mathfrak b}}^{(s+1)}_{0}.
\eeq

{\bf Step 4.} Having all the polynomials $P_{(s,k)}(g)$ for $s,k\le r$ determined, we define the general solution for
${\mathfrak b}^{(g)}_{g-r-1}$:
\beq
\label{solution-bgr}
{\mathfrak b}^{(g)}_{g-r-1}=\sum_{s=0}^r P_{(s,r-s)}(g)\prod_{l=1}^{g-1-r}q(s+1+l,s){\wht{\mathfrak b}}^{(s+1)}_{0},
\eeq
where the coefficients ${\wht{\mathfrak b}}^{(s+1)}_{0}$ are to be defined from the triangular system of
linear equations originating from the fact that ${\mathfrak b}^{(g)}_{0}=1/(2g+1)$:
\beq
{\wht{\mathfrak b}}^{(1)}_{0}=\frac13,\quad
{\wht{\mathfrak b}}^{(s+1)}_{0}=\frac1{2s+3}-\sum_{p=0}^{s-1}P_{(p,s-p)}(s+1){\wht{\mathfrak b}}^{(p+1)}_{0}, \ s>0.
\eeq

We can therefore derive the coefficients $\varkappa_{g,1,r}$ for any $g$ for a fixed ``degree of reduction'' $r$.

\begin{remark}\label{rem-Psk}
The first few polynomials $P_{(s,k)}(g)$ are
\bea
&{}& P_{(s,0)}(g)=1,\nonumber\\
&{}& P_{(s,1)}(g)=\frac15[3g-s-3],\nonumber\\
&{}& P_{(s,2)}(g)=\frac{9}{50}g^2-\frac{33}{70}g-\frac{3}{25}gs+\frac{1}{50}s^2+\frac{11}{35}s+\frac{39}{175}.
\nonumber
\eea
So, we can conclude that the quantities $P_{(s,k)}(g)$ are polynomials of degree $k$ in both $g$ and $s$, and
their general form is
\beq
P_{(s,k)}(g)=\sum_{{i\ge 0,\ j\ge 0\atop i+j\le k}}a^{(k)}_{i,j}g^is^j.
\label{Psk}
\eeq
From relation (\ref{rec-P-exp}) we can read the recurrence relations on the coefficients $a^{(k)}_{i,j}$:
rewriting (\ref{rec-P-exp}) in the form
\bea
&{}&(3g-3-k-s)\left[\sum_{i+j\le k}a^{(k)}_{i,j}\bigl( (g+1)^i-g^i\bigr)s^j
-\sum_{i+j\le k-1}a^{(k-1)}_{i,j}g^is^j\right]\nonumber\\
&{}&\qquad\qquad =-2\sum_{i+j\le k-1}a^{(k-1)}_{i,j}g^is^j-2k \sum_{i+j\le k}a^{(k)}_{i,j} g^is^j,
\label{aij}
\eea
which produces the recurrence relations if we equate coefficients of $g^i s^j$:
\begin{itemize}
\item for $i+j\le k-1$,
\bea
&{}&3\sum_{p=i}^{k-j}\binom{i-1}{p} a^{(k)}_{p,j}+(3-k)\sum_{p=i+1}^{k-j}\binom{i}{p}a^{(k)}_{p,j}
-\sum_{p=i+1}^{k+1-j}\binom{i}{p}a^{(k)}_{p,j-1}-3a^{(k-1)}_{i-1,j}-(3-k)a^{(k-1)}_{i,j}+a^{(k-1)}_{i,j-1}
\nonumber\\
&{}&\qquad\qquad =-2a^{(k-1)}_{i,j}-2k a^{(k)}_{i,j},
\label{aij1}
\eea
where the terms with $i-1$ are absent for $i=0$ and those with $j-1$ are absent for $j=0$;
\item $i+j=k$, \ $i>0$, $j>0$:
\beq
3i a^{(k)}_{i,j}-(i+1)a^{(k)}_{i+1,j-1}-3a^{(k-1)}_{i-1,j}+a^{(k-1)}_{i,j-1}+2k a^{(k)}_{i,j}=0;
\label{aij2}
\eeq
\item $i=0$, $j=k$:
\beq
-a^{(k)}_{1,k-1}+a^{(k-1)}_{0,k-1}+2k a^{(k)}_{0,k}=0;
\label{aij3}
\eeq
\item $i=k$, $j=0$:
\beq
5k a^{(k)}_{k,0}=3a^{(k-1)}_{k-1,0},\quad\hbox{or}\quad a^{(k)}_{k,0}=\frac{3^k}{5^k\, k!}.
\label{aij4}
\eeq
\end{itemize}

We see that the quantities $a^{(k)}_{i,j}$ with $i+j=k$ satisfy the separate recursion relations
(\ref{aij2})--(\ref{aij4}); we can first solve these relations, determining subsequently
all lower degree terms from the relation (\ref{aij1}).
\end{remark}

\subsection{Deriving $b^{(g)}_{g-2}$ from the graph representation of Lemma \ref{lm:graph}}

We now derive the first subleading coefficient $b^{(g)}_{g-2}$ (the first ancestor invariant)
of the expansion of the one-loop mean $L_g(e^{\lambda}+e^{-\lambda})$ (\ref{general}) using 
the decomposition formulas and Lemma~\ref{lm:graph}.

We consider only the part with the times $T^+_{2k}$. The highest term for genus $g$ is $\<\tau_{3g-2}\>_gT^+_{6g-4}$

\begin{figure}[h]
{\psset{unit=1}
\begin{pspicture}(-5,-2)(3,2)
\newcommand{\PATTERN}{%
{\psset{unit=1}
\pscircle[fillstyle=solid,fillcolor=gray](0,0){1}
\psline[linewidth=2pt,linecolor=blue]{->}(0,-0.8)(0,-1.5)
\rput(0,-1.5){\makebox(0,0)[ct]{$T^+_{6g-6}$}}
}
}
\rput(-2,0){\PATTERN}
\rput(-2,0){\makebox(0,0)[cc]{$g{-}1$}}
\psarc[linecolor=red,linestyle=dashed,linewidth=2pt](-2,1){0.8}{-30}{210}
\rput(-2.6,0.55){\makebox(0,0)[lc]{$0^+$}}
\rput(-1.4,0.55){\makebox(0,0)[rc]{$0^+$}}
\rput(0,0){\makebox(0,0)[cc]{$+$}}
\rput(2,0){\PATTERN}
\rput(2,0){\makebox(0,0)[cc]{$g$}}
\psline[linewidth=2pt,linecolor=blue]{-|}(2,0.8)(2,1.3)
\rput(2,1.4){\makebox(0,0)[cb]{$2^+$}}
\end{pspicture} }
\caption{\small The two diagrams contributing to the first-order correction to
the Kontsevich potential: the external edge carries the time $T^+_{6g-6}$, the first
graph contains one internal edge with the labeling $(0,0)$ whereas the second graph contains one half-edge
with the labeling $2$.}
\label{fi:firstcorrection}
\end{figure}
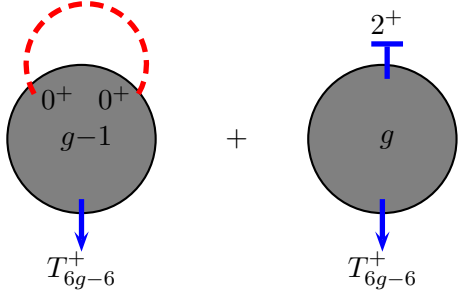

Following Lemma~\ref{lm:graph}, the first-order
correction, or the coefficient of $T^+_{6g-6}$, comes only from two terms: from the graph with
one vertex and one internal edge with endpoint markings $(0,0)$ and from the graph with one vertex and one half-edge
with marking 2 (see Fig.~\ref{fi:firstcorrection}): the corresponding coefficient is then
\beq
\frac{B_2}{4}\<\tau_{3g-3}\tau_0\tau_0\>_{g-1}+\frac{2^3}{5!}\<\tau_{3g-3}\tau_2\>_g,
\label{bg-g-2-rec}
\eeq
and we need only to know the corresponding intersection indices. Whereas
$\<\tau_{3g-3}\tau_0\tau_0\>_{g-1}=\<\tau_{3g-5}\>_{g-1}$, the intersection index
$\<\tau_{3g-3}\tau_2\>_g$ must be calculated separately, and we do it in Lemma~\ref{lm:index} in
Appendix~\ref{Ap:tau}. Using formula (\ref{index}) and that $B_2=1/24$, we obtain that the
coefficient in front of $T^{+}_{6g-6}$ is
\beq
\frac15 [12g^2-7g+5]
\label{coefficient}
\eeq
in full agreement with (\ref{c-g-1-1}).

\subsection*{Conclusion}
Application of topological recursion (TR) to constructing generating 
functions for cohomological field theories is becoming an important issue in 
contemporary mathematical physics (see, e.g., the recent paper \cite{FLZ14} where all genus all descendants equivariant Gromov-Witten invariants of ${\mathbb P}^1$ were constructed using TR. In this respect, it seems interesting to understand the status of 
Givental-type decompositions in the quantum spectral curve approach.

\subsection*{Acknowledgments}
This paper has been started in 2011 and during the years of writing it we had numerous 
useful discussions with B. Eynard, A.D. Mironov, S. Shadrin, P. Dunin-Barkovsky, M. Mulase, 
C. Reidys, P. Su{\l}kowski, and Don Zagier to all of whom we are deeply grateful. 

The work of J.E.A., L.O.Ch. and R.C.P. was partially supported by the center of excellence grant ``Centre for Quantum Geometry of Moduli Spaces'' from the Danish National Research Foundation (DNRF95). The work of L.O.Ch. was supported by the Russian Foundation for Basic Research (Grant Nos. 14-01-00860-a and 13-01-12405-ofi-m2). The work of P.N. was partially supported by the Australian Research Council grant DP1094328.

\def\thetheorem{\Alph{section}.\arabic{theorem}}
\def\theprop{\Alph{section}.\arabic{prop}}
\def\thelemma{\Alph{section}.\arabic{lemma}}
\def\thecor{\Alph{section}.\arabic{cor}}
\def\theexam{\Alph{section}.\arabic{exam}}
\def\theremark{\Alph{section}.\arabic{remark}}
\def\theequation{\Alph{section}.\arabic{equation}}

\setcounter{section}{0}

\appendix{Calculating $\<\tau_{3g-3}\tau_2\>_g$}\label{Ap:tau}
\setcounter{equation}{0}

The Kontsevich model partition function ${\mathcal Z}_{\text{K}}=e^{{\mathcal F}_{\text{K}}[T_{2k}]}$ satisfies the set of Virasoro conditions:
\be
L_{n} {\mathcal Z}_{\text{K}}=0,\quad n=-1,0,1,\dots ,
\ee 
where 
\bea
L_n&=&(2n+3)!!\frac{\pa}{\pa T_{2(n+1)}} +\sum_{i=0}^\infty \frac{(2i+2n+1)!!}{(2i-1)!!}
T_{2i}\frac{\pa}{\pa T_{2(i+n)}}\nonumber\\
&{}&+\frac12 \sum_{i=0}^{n-1}(2n-2i-1)!!(2i+1)!!\frac{\pa^2}{\pa T_{2i}\pa T_{2(n-i-1)}}
+\delta_{n,-1}T_0^2/2+\delta_{n,0}1/16.
\label{Virasoro}
\eea
These conditions for $n=-1$ and $n=0$ are the corresponding string and dilaton equations which give
conditions on lower $\psi$-classes:
\bea
&{}&\Bigl\langle \tau_0 \prod_{i=1}^s \tau_{r_i}\Bigr\rangle_g
=\sum_{j=1}^s \Bigl\langle \prod_{i=1}^s \tau_{r_i-\delta_{ij}}\Bigr\rangle_g\label{string1}\\
&{}&\Bigl\langle \tau_1 \prod_{i=1}^s \tau_{r_i}\Bigr\rangle_g
=(2g-2+s) \Bigl\langle \prod_{i=1}^s \tau_{r_i}\Bigr\rangle_g.\label{dilaton}
\eea
For $L_1$ we have
\be
L_1=-3\cdot 5 \frac{\pa}{\pa T_4}+\sum_{i=0}^\infty (2i+3)(2i+1) T_{2i}\frac{\pa}{\pa T_{2i+2}}+\frac12\frac{\pa^2}{\pa T_0^2}
\ee
and gathering the coefficients of the linear term in $T_{6g-6}$, we obtain
\bea
3\cdot 5 \langle \tau_2\tau_{3g-3}\rangle_g&=&(6g-3)(6g-5)\langle \tau_{3g-2}\rangle_g
+\frac12 \langle \tau_0^2 \tau_{3g-3}\rangle_{g-1}\nonumber\\
&=&\bigl[3(2g-1)(6g-5)+24g/2\bigr]\langle \tau_{3g-2}\rangle_g=3[12g^2-12g+5]\langle \tau_{3g-2}\rangle_g,\nonumber
\eea

We therefore obtain the following technical lemma
\begin{lemma}\label{lm:index}
For the intersection indices, we obtain
\beq
\<\tau_2\tau_{3g-3}\>_{g}=\frac15 [12g(g-1)+5]\<\tau_{3g-2}\>_g,\quad g\ge2.
\label{index}
\eeq
\end{lemma}

\end{document}